\documentclass{emulateapj}
\usepackage{graphics}
\usepackage{epsf}
\usepackage{amsmath}
\usepackage{natbib}
\usepackage[colorlinks=true,  citecolor=blue,
  linkcolor=blue,  menucolor=blue, urlcolor=blue,
  linkbordercolor={0 0 1}, frenchlinks=True, breaklinks]{hyperref}

\newcommand{\Mstar}{$M_{\star}$}
\newcommand{\MZ}{\Mstar--$Z$}
\newcommand{\MSFR}{\Mstar--SFR}
\newcommand{\MZS}{\MZ--SFR}
\newcommand{\OII}{[{\rm O}~\textsc{ii}]}
\newcommand{\OIII}{[{\rm O}~\textsc{iii}]}
\newcommand{\NII}{[{\rm N}~\textsc{ii}]}

\newcommand{\Msun}{$M_{\odot}$}
\newcommand{\GALEX}{{\it GALEX}}
\newcommand{\SFRHa}{{\rm SFR}_{{\rm H}\alpha}}
\def \Z {$Z$}
\def \Ha {H$\alpha$}
\def \Hb {H$\beta$}
\begin{document}

\title{The Relationship between Stellar Mass, Gas Metallicity, and Star
  Formation Rate for H$\alpha$-selected Galaxies at $z\approx0.8$ from
  the NewH$\alpha$ Survey}
\author{
Mithi A. de los Reyes\altaffilmark{1,2}, 
Chun Ly\altaffilmark{2,3,10}, 
Janice C. Lee\altaffilmark{2,4}, 
Samir Salim\altaffilmark{5}, 
Molly S. Peeples\altaffilmark{2},
Ivelina Momcheva\altaffilmark{6}, 
Jesse Feddersen\altaffilmark{6,2}, 
Daniel A. Dale\altaffilmark{7}, 
Masami Ouchi\altaffilmark{8}, 
Yoshiaki Ono\altaffilmark{8}, and 
Rose Finn\altaffilmark{9}
}
\email{madelosr@ncsu.edu}
\altaffiltext{1}{Department of Physics, North Carolina State University, Raleigh, NC}
\altaffiltext{2}{Space Telescope Science Institute, Baltimore, MD}
\altaffiltext{3}{National Aeronautics and Space Administration, 
  Goddard Space Flight Center, Greenbelt, MD, USA}
\altaffiltext{4}{Visiting Astronomer, SSC/IPAC, Caltech, Pasadena, CA}
\altaffiltext{5}{Astronomy Department, Indiana University, Bloomington, IN}
\altaffiltext{6}{Astronomy Department, Yale University, New Haven, CT}
\altaffiltext{7}{Department of Physics and Astronomy, University of Wyoming, Laramie, WY}
\altaffiltext{8}{Institute for the Physics and Mathematics of the Universe, TODIAS,
  University of Tokyo, Japan}
\altaffiltext{9}{Physics Department of Physics and Astronomy, Siena College, Loudonville, NY}
\altaffiltext{10}{NASA Postdoctoral Fellow.}
\date{\today}

\defcitealias{cha03}{Chabrier}
\defcitealias{m91}{M91}
\defcitealias{kk04}{KK04}
\defcitealias{lar10}{Lar10}
\defcitealias{lar13}{Lar13}
\defcitealias{man10}{Man10}
\defcitealias{tre04}{T04}
\defcitealias{yat12}{Yat12}
\defcitealias{z94}{Z94}

\newcommand{\Nspec}{299}	   
\newcommand{\Nagn}{21}		   
\newcommand{\Nnoagn}{278}	   
\newcommand{\Nfiveagn}{160}	   
\newcommand{\Nbadfiveagn}{118}	   
\newcommand{\percentfiveagn}{57.6} 
\newcommand{\Nthreeagn}{255}	   
\newcommand{\Nbadthreeagn}{23}	   
\newcommand{\Nthreenew}{274}	   
\newcommand{\Nbadthreenew}{25}	   
\newcommand{\Nfivenew}{174}	   
\newcommand{\Nbadfivenew}{125}	   
\newcommand{\Nonefournew}{201}	   
\newcommand{\Nbadonefournew}{62}   
\newcommand{\Nfivesignew}{137}	   
\newcommand{\Nthreesignew}{187}	   
\newcommand{\NSFRcompare}{114}	   
\newcommand{\Nmzfivesig}{98}	   
\newcommand{\Nthree}{119}	   

\begin{abstract}
Using a sample of \Nspec\ H$\alpha$-selected galaxies at $z\approx0.8$, we study the 
relationship between galaxy stellar mass, gas-phase metallicity, and star formation 
rate (SFR), and compare to previous results. We use deep optical spectra obtained 
with the IMACS spectrograph at the Magellan telescope to measure strong oxygen lines.
We combine these spectra and metallicities with (1) rest-frame UV-to-optical 
imaging, which allows us to determine stellar masses and dust attenuation corrections, 
and (2) H$\alpha$ narrowband imaging, which provides a robust measure of the 
instantaneous SFR. Our sample spans stellar masses of $\sim$10$^9$ to $6\times10^{11}$ 
$M_{\sun}$, SFRs of 0.4 to 270 $M_{\sun}$ yr$^{-1}$, and metal abundances of
$12+\log({\rm O/H})\approx8.3$--9.1 ($\approx0.4$--2.6 $Z_{\sun}$). 
The correlations that we find between the H$\alpha$-based SFR and stellar mass
(i.e., the star-forming ``main sequence''), and between the stellar mass and
metallicity, are both consistent with previous $z\sim1$ studies of star-forming galaxies.
We then study the relationship between the three properties using various plane-fitting 
techniques (Lara-L\'opez et al.) and a curve-fitting projection (Mannucci et al.). 
In all cases, we exclude strong dependence of the \MZ\ relation on SFR, but are unable
to distinguish between moderate and no dependence. Our results are consistent with previous 
mass-metallicity-SFR studies. We check whether dataset limitations may obscure a strong 
dependence on the SFR by using mock samples drawn from the SDSS. These experiments reveal 
that the adopted signal-to-noise cuts may have a significant effect on the measured 
dependence. Further work is needed to investigate these results, and to test whether a 
``fundamental metallicity relation'' or a ``fundamental plane'' describes star-forming
galaxies across cosmic time.
\end{abstract}

\section{Introduction}
Studying the general relationships between the physical properties of galaxies---including
stellar mass (\Mstar), gas-phase metallicity (\Z), and star formation rate (SFR)---provides
clues about galaxy formation and evolution. Stellar mass is an estimate of the amount of
gas converted into stars in a galaxy over time, while the SFR measures the current rate at
which gas is consumed to form stars. In addition, the gas-phase metallicity reflects both
the amount of gas reprocessed by stars and galactic interactions with the environment
through the infall and outflow of gas.

Combinations of these three properties have been well-studied. The mass--metallicity (\MZ)
relation is a nonlinear one in which \Z\ increases with \Mstar\ up to a stellar mass of
about $3\times10^{10}$ \Msun\ and then plateaus (e.g., \citealt{tre04}, hereafter T04;
\citealt{mou11,zah11,and13}).
The relation has been shown to evolve towards lower metallicity at higher redshifts
\citep{man09,erb06,zah13a,mai08}, although the exact nature of this evolution is unclear,
in part because high-$z$ results are still significantly incomplete at low stellar masses.
Similarly, the positive correlation between \Mstar\ and SFR
\citep[SFR $\propto M_{\star}^{0.6}$;][]{sal12}; called the ``star-formation sequence''
\citep{sal07} or the galaxy ``main sequence'' \citep{noe07} shows evolution with
redshift toward higher SFRs at earlier times \citep[e.g.,][]{noe07,elb07,whi12}.
  
Despite the tightness of the \MZ\ relation, some intrinsic scatter remains ($\sim$0.1 dex).
It has been suggested that part of this scatter can be accounted for by a secondary
dependence on the SFR---at a given stellar mass, lower-metallicity galaxies tend to have
higher SFRs (\citealp{ell08,man10,lar10}), and lower SFR galaxies tend to have higher
metallicities (\citealp{ell08, pee08}). A relationship between all three properties was
proposed \citep[e.g.,][hereafter Man10]{man10}. 

The physical origin of such a \MZS\ relation is thought to be a result of the way galaxies
process gas. The oxygen-to-hydrogen\footnote{Oxygen is generally used as a proxy for total
  metal content, since it is the most abundant gas-phase metal in a galaxy.}
gas ratio in a galaxy is regulated by its stellar mass, history of outflows, and gas mass.
The amount of oxygen in the interstellar medium (ISM) is primarily set by the mass
of oxygen the galaxy has produced in its lifetime (which is roughly proportional to its
stellar mass), less the oxygen mass residing in stars and/or lost in outflows \citep{pee14}.
The degree to which the ISM oxygen content is diluted is determined by the galaxy's gas
mass \citep{pee11}, which is in turn regulated by a balance between star
formation, accretion, and outflows \citep[e.g.,][]{dav11,lil13,for13}. 

Infall of metal-poor gas will initially dilute the metal abundance already present in
the ISM while enhancing the SFR, leading to the observed trend of $Z$ and SFR inversely
proportional at a given mass. However, as the enhanced star formation continues, the
freshly produced metals can quickly erase the signature of fresh inflow, causing an
{\em increase} in metallicity while the SFR is still relatively high \citep{tor12}. 
Outflows driven by star formation must then be removing these freshly produced metals
from the ISM in order for the galaxy to continue to have a low ISM abundance in the
while having enhanced star formation. In this scenario, an observed \MZS\ relation
(and its assumed lack of evolution) is largely coincidental and a result of the tendency
for galaxies to move toward an equilibrium between galactic inflows and outflows
\citep{dav11}.

In this framework, if galaxies at different redshifts universally obey the same relation
between stellar mass, gas mass, and gas metallicity, then it could imply something
``fundamental'' about how galaxies expel their metals through time. Measurements of
\textsc{H i} masses are not currently feasible at redshifts beyond the local universe,
so SFR has generally been used as a proxy for gas mass. Furthermore, a ``fundamental''
\MZS\ relation would imply that the evolution of the \MZ\ relation and the star-formation
sequence are simply consequences of preferentially observing higher SFRs at higher
redshifts (\citetalias{man10}; \citealp{hun12}). 

However, the existence of a fundamental \MZS\ relation remains controversial.
While several works have found evidence of such a relation at local redshifts
(see e.g., \citetalias{man10}; \citealp[][hereafter Yat12]{hun12,and13,per13,yat12}),
\citet{san13} and \citet{hug13} were unable to find a significant dependence of the
\MZ\ relation on the total SFR from integral field spectroscopy at local redshifts,
and suggested that previously reported results may be due to the impact of
observational effects such as aperture bias on the SFR.
At higher redshifts, uncertainty remains over the existence of the \MZS\ relation and
its evolution. Once again, several studies found evidence of a relation
\citep[e.g.,][whose results all agreed with the local \MZS\ relation]{ric11,cre12,bel13,hen13a,hen13b},
but other studies were less conclusive. \citet{yab12} and \citet{yab14} found a \MZS\
relation deviating slightly from that reported at local redshifts, while \citet{zah13b}
found a weak dependence of the \MZ\ relation on SFR that was significantly different 
from the local relation and concluded that this was a result of redshift evolution.
In addition, \citet{sto13} stacked spectra and found evidence that star-forming galaxies
at $z\sim0.8$ and $z\sim1.5$ have gas-phase metallicities that are consistent with the
local \MZ\ relation, in contrast with other high-$z$ studies.
In most cases, samples are potentially subject to dataset limitations.

Ideally, tracking \Mstar, \Z, and SFR in a consistent manner across a range of redshifts
would provide a solid empirical basis from which to study their relationship and its
evolution. At low-$z$, the \MZS\ relation has largely been investigated using data from
the Sloan Digital Sky Survey \citep[SDSS;][]{yor00}, a sample consisting of over a hundred
thousand galaxies that covers stellar masses from about $10^{9}$ to $10^{11.5}$ \Msun,
gas-phase metallicities from $12+\log{({\rm O/H})} = 8.5$ to 9 (0.6--2 $Z_{\odot}$), and SFRs
from $\log({\rm SFR}/(M_{\odot}~{\rm yr}^{-1}))=-1.45$ to 0.8 \citepalias{man10}.
The \MZS\ relation has also been extended to low-mass galaxies, albeit with smaller sample
sizes, down to \Mstar\ $\sim10^{8.3}$ \Msun\ with gamma-ray bursts \citep{man11} and
$\sim10^{6}$ \Msun\ with dwarf galaxies \citep{hun12}. 

Studies at higher redshifts generally have much smaller samples that cover more limited
portions of parameter space. For $z=0.4$--1, \citetalias{man10} used a sample of 69
galaxies with masses of $10^{8.2}$--$10^{10.7}$ \Msun\ from \citet{sav05}, while
\citet{lar10} (hereafter Lar10) used 88 galaxies from \cite{rod08} with masses of
$10^{9}$--$10^{11.2}$ \Msun.
At $1\lesssim z\lesssim 3$, the largest samples used are those of \cite{erb06} with 91
UV-selected galaxies at $z\sim2.2$, and that of the Spectroscopic Imaging Survey
\citep{for09}, consisting of 62 galaxies at $z\sim2$, both with mass ranges of
$\sim$10$^{9}$--$10^{11}$ \Msun. 
Recent efforts have increased the sample size at $z\sim1.5$. \citet{zah13b} used
$\sim$150 star-forming galaxies from the COSMOS field at $z\sim1.6$ and masses
ranging from approximately $10^{9.5}$ to $10^{11}$ \Msun.
\citet{yab12} and \citet{yab14} conducted near-infrared fiber spectroscopy for 70--340
galaxies at $z\sim1.4$, \citet{sto13} used 64 \Ha-selected galaxies at $z\sim1$--1.5,
and \citet{hen13b} performed infrared grism spectroscopy for 83 galaxies at
$z\sim1.5$--2.3.
However, at high-$z$, spectral stacking is predominantly used, and low-mass galaxies
below $5\times10^9$ \Msun\ have yet to be studied extensively.
There have been efforts to extend $z\approx0.5$--3 studies toward lower stellar masses
\citep{xia12,ly14,hen13a,hen13b,bel13}; however, sample sizes remain limited.

Thus, the \MZS\ relation requires further study, particularly for redshifts above
$z\sim0.3$ (the maximum redshift for the SDSS sample). We aim to build upon previous
work by using a sample of star-forming galaxies at $z=0.8$ (when the universe was
roughly $\sim7$ Gyr old or half of the Hubble time). 
We use methods of deriving \Mstar, $Z$ and SFR which are similar to those used by
local studies. Deep rest-frame optical spectra obtained with the IMACS spectrograph at
the Magellan 6.5-m telescope are used to measure gas-phase metallicities with oxygen
strong-line calibrations. The spectra are used along with (1) rest-frame UV-to-optical
imaging data, which allow us to determine stellar masses and dust attenuation
corrections, and (2) our H$\alpha$ narrowband imaging data, which provide a robust
measure of SFR. 

In Section \ref{sec:data}, we describe the NewH$\alpha$ survey, sample selection,
and spectroscopy. We also present the photometric properties and the spectroscopic
emission-line fluxes of the galaxies used in this analysis.
Section \ref{sec:prop} discusses the calculation of our physical properties from (1)
spectral energy distribution (SED) fitting spanning rest-frame 1400--7000 \AA\ to
estimate stellar masses, dust attenuation and UV SFRs, (2) H$\alpha$ luminosities to
determine SFRs, and (3) several empirical and theoretical strong-line calibrations to
estimate gas-phase metallicity. In Section \ref{sec:analysis}, we use our data to
produce a \MZ\ relation and a \MSFR\ relation, and compare them with previous
literature results. We also investigate the \MZS\ relation through different
plane-fitting and three-dimensional curve-fitting approaches and compare with
literature \MZS\ relations found at local redshifts.
In Section~\ref{sec:disc}, we discuss how limitations in: (1) our dataset, (2)
plane-fitting techniques, and (3) the parameterization of the \MZS\ relation may
affect our ability to fully constrain the existence or evolution of the \MZS\
relation. Finally, we summarize our work in Section \ref{sec:concl}.

Throughout this paper, we assume a $\Lambda$CDM cosmology with $H_0$ = 70 km s$^{-1}$
Mpc$^{-1}$, $\Omega_M=0.3$, and $\Omega_{\Lambda}=0.7$ for distance-dependent calculations,
which is similar to the Seven-Year WMAP results \citep{kom11}. Unless otherwise noted,
a \cite{cha03} initial mass function (IMF; hereafter Chabrier) is assumed, and all
magnitudes are reported on the AB system \citep{oke74}.
 
\section{Data}
\label{sec:data}
We present near-IR narrowband photometry and optical spectroscopy for a sample of
\Nspec\ H$\alpha$-selected galaxies at $z\sim0.8$ from the NOAO Extremely
Wide-Field Infrared Imager (NEWFIRM) NewH$\alpha$ survey \citep{ly11a,lee12}.
The galaxies are identified in the Subaru-XMM Deep Survey (SXDS) field \citep{fur08},
and deep follow-up spectroscopy was performed with the Inamori Magellan Areal Camera
and Spectrograph \citep[IMACS;][]{dre06} at the Magellan 6.5-m Baade telescope
\citep{mom13}.
Emission lines from \OII\,$\lambda$3727 to \OIII\,$\lambda$5007 are observed in the
spectra. In this section, we provide a brief overview of the NewH$\alpha$ survey, the
sample selection, and the IMACS spectroscopy.  We also describe multi-wavelength
broadband photometry used in SED fitting to derive galaxy stellar masses, dust
attenuation, and a second measure of the dust-corrected SFR.

\subsection{The NewH$\alpha$ Narrowband Survey: Selecting Emission-Line Galaxies}

The NewH$\alpha$ Survey is a program that has obtained emission-line selected samples
at intermediate redshift \citep{lee12}.  The program was designed to efficiently obtain
statistical samples of both luminous (but rare) and faint emission-line galaxies. We
do this by combining the near-IR imaging capabilities of NEWFIRM \citep{aut03,pro08}
at the KPNO/CTIO 4-m telescopes to cover large areas (field-of-view of 27\farcm6
$\times$ 27\farcm6), and FourStar \citep[field-of-view of 10\farcm9 $\times$
10\farcm9;][]{per08} at the Magellan 6.5-m Baade telescope to probe luminosities that
are about a factor of three deeper over smaller areas. For both cameras, we use a pair
of custom 1\% narrowband filters that fit within windows of high atmospheric
transmission and low OH airglow at 1.18 $\mu$m and 2.09 $\mu$m. With these two filters,
deep H$\alpha$-selected galaxy samples are obtained at $z=0.8$ (near the beginning of
the ten-fold decline in the cosmic SFR density) and at $z=2.2$
\citep[near the peak of the cosmic SFR density; see, e.g.,][]{red08,ly09,ly11b}.

\begin{figure}
  \epsscale{1.1}
  \plotone{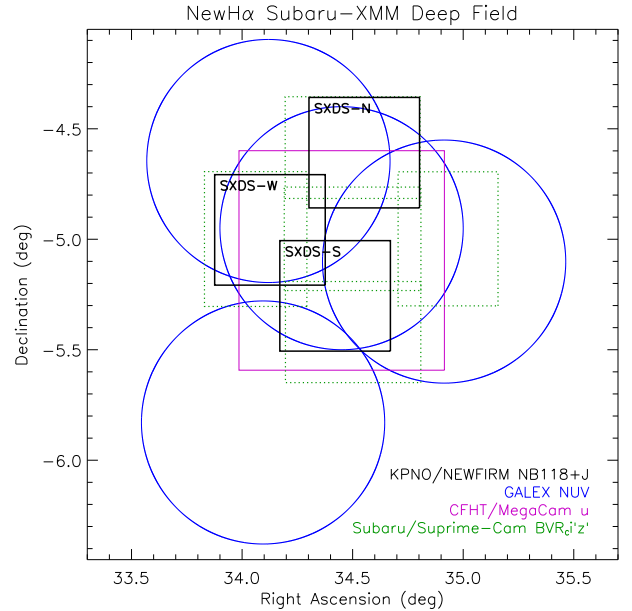}
  \caption{New\Ha\ and multi-wavelength ancillary data for the SXDS field.
    Our NEWFIRM imaging data are shown by labeled squares 
    for three pointings (SXDS-N, SXDS-S, and SXDS-W), along with \GALEX\ NUV
    (circles, blue in online version), 
    Subaru/Suprime-Cam $BVR_{\rm C}i\arcmin z\arcmin$ (dashed 
    rectangles, green in online version), and CFHT/Megacam 
    $u$-band imaging (square, purple in online version).
    The latter also illustrates the imaging footprint for the UKIRT's
    Ultra Deep Survey in $JHK$.}
  \label{fig:1}
\end{figure}

The work presented in this paper focuses on H$\alpha$ emitters at $z=0.8$, which are
detected in NEWFIRM narrowband 1.18$\mu$m (hereafter NB118) and $J$ imaging of a 0.82
deg$^2$ region in the SXDS (see Figure \ref{fig:1}). Here we summarize the NB118
observations, data reduction, and selection method used to produce samples of
emission-line galaxy candidates that are then targeted for IMACS spectroscopy.

NEWFIRM at the KPNO 4-m was used to obtain observations over three pointings in the
SXDS ($\alpha$ = 2$^h$18$^m$; $\delta$ = --5\arcdeg) in 2007 December, 2008 September,
and 2008 October.  The positioning of these three fields relative to other observations
are shown in Figure \ref{fig:1}. The cumulative exposure times for each pointing
ranged from 8.47 to 12.67 hr in NB118 and from 2.40 to 3.97 hr in $J$. The median
seeing during our observations was ~1\farcs2, and varied between 1\farcs0 and 1\farcs9,
so point sources are adequately sampled by NEWFIRM's 0\farcs4 pixels. Standard near-IR
deep-field observing procedures and reduction techniques were used, and are discussed
further in \cite{ly11a}. The 3$\sigma$ limiting magnitudes, in apertures (of diameters
twice the FWHM) containing at least $\sim$80\% of the flux of a point source, range
from 23.7 to 24.2 mag (23.4 to 24.1 mag) in NB118 ($J$).

Sources are selected as emission-line galaxy candidates if they show a $J$--NB118 color
excess that is significant at the 3$\sigma$ level and is greater than 0.2 mag. The
minimum of 0.2 mag is based on the scatter in the color excess for bright point sources.
Corrections for the continuum slope are applied based on the $z^{\prime}$--$J$ color
\citep{ly11a}, using publicly available Subaru/Suprime-Cam $z^{\prime}$ data where
available (see Section~\ref{sec:phot} for further details). The overall procedure
follows general selection techniques commonly used in narrowband surveys
\citep{fuj03, ly07, shi08, vil08, sob09}. A total sample of 661 emission-line galaxy
candidates meeting these criteria was obtained over the three NEWFIRM pointings.
Follow-up spectroscopy was obtained for a subset of these galaxies, as described below.
Using a combination of color-selection methods and spectroscopic confirmation,
approximately half of these candidates are identified as H$\alpha$ excess emitters
at $z\approx0.8$ \citep{ly11a}.

\subsection{IMACS Spectroscopy}
\label{sec:imacs}
As discussed in \citet{mom13}, deep follow-up spectroscopy of the NewH$\alpha$ NB118
emission-line galaxy candidate sample was performed in 2008--2009 with IMACS on the
Magellan-I telescope. IMACS enables multi-object spectroscopy with slit-masks over
a 27\farcm4 diameter area (well-matched to NEWFIRM's field-of-view), and has good
sensitivity to $\sim$9500 \AA. These two characteristics make IMACS an ideal instrument
for optical spectroscopic follow-up of NewH$\alpha$ NB118 excess sources, and in
particular, H$\alpha$ emitters at $z\approx0.8$.

The chosen observational setup yields spectral coverage from 6300 \AA\ to 9600 \AA\
(corresponding to rest-frame $\sim$3500 \AA\ to 5300 \AA), and captures the strong
rest-frame optical emission lines from \OII\,$\lambda$3727 (observed at $\approx$6700
\AA) to \OIII\,$\lambda$5007 (observed at $\approx$9000 \AA) for galaxies at
$z\approx0.8$.  Slit widths of 1\farcs5 were chosen ($\sim$11 kpc at $z=0.8$). The
seeing during our observing runs was generally sub-arcsecond.
The typical integration time was 4.5 hours; however, for about half of the \Ha-emitting
galaxies, deeper observations were acquired for a total of 7.75 hours
\citep[see Figure 1 of][]{mom13} to improve the measurement of detected, but low
signal-to-noise (S/N) Balmer lines.

Of the 661 NB118 emission-line galaxy candidates in the SXDS, 386 were targeted with
IMACS.  Priority was given to sources likely to be intermediate redshift candidates
based on their photometric redshifts \citep{fur08}, while galaxies with low photometric
redshift ($z_{\rm phot}<0.7$) were used as slit-mask fillers.  Sources which showed an
NB118 excess at a significance lower than the 3$\sigma$ cutoff were also used as fillers.

Overall, 225 (74) of the 3$\sigma$ ($<$3$\sigma$) selected sample observed with IMACS
have spectroscopic redshift ($z_{\rm spec}$) between 0.78 and 0.83, confirming that the
narrowband photometric excess is due to H$\alpha$ emission.
Note that the redshift range about $z=0.8$ is slightly larger than expected. This is
because not all light entering the narrowband filter is normally incident to the
filter. As the angle of incidence increases, redshifts are biased blueward, increasing
the expected redshift range.

Standard spectroscopic multi-object observational techniques were followed, and
long-slit observations of spectrophotometric standards were obtained for flux
calibration. Data reduction was performed using the dedicated software package called
``COSMOS'', developed by the IMACS instrument team at Carnegie
Observatories.\footnote{\url{http://code.obs.carnegiescience.edu/cosmos}.}
To obtain continuum-subtracted and absorption-corrected line flux measurements, the
fluxed spectra were fit with stellar population models. This fitting method, similar
to that used for the SDSS (\citetalias{tre04}; \citealt{bri04}), assumes that any
galaxy star formation history (SFH) can be approximated by a sum of discrete bursts
(simple stellar populations). Foreground extinction was corrected using the
\cite{sch98} extinction map and the \cite{odo94} Milky Way extinction curve; for
SXDS, average $E(B-V)=0.03$.

In Table \ref{table:genprop}, we present the NEWFIRM photometric properties of the
\Nspec\ spectroscopically confirmed H$\alpha$ emitters in the SXDS with IMACS. That
is, these galaxies have $0.78<z_{\rm spec}<0.83$. These data are used to compute the
\Ha-based SFRs used in the analysis, as discussed in Section \ref{sec:SFR_comp}.
Table \ref{table:emline_flux} gives the IMACS spectroscopic fluxes for the strong
rest-frame optical oxygen emission lines, as well as the Balmer lines (H$\beta$,
H$\gamma$, and H$\delta$). Active galactic nuclei (AGN) classification was performed
by \citet{mom13} using a combination of methods, including UV variability, line
widths, and emission-line diagnostics with the ``Mass-Excitation'' diagram
\citep{jun11}, to identify both broad- and narrow-line AGN. The \Nagn\ galaxies
determined to have AGN are identified in Table \ref{table:msfr_agn}, and are excluded
from the remainder of the analysis, leaving a working sample of \Nnoagn. For the
remaining galaxies, the line fluxes are used to compute the gas-phase metallicities,
as discussed in Section \ref{sec:metal}.

\subsection{Multiwavelength Photometry}
\label{sec:phot}
We use multiband photometry (NUV, $u$, $B$, $V$, $R_{\rm C}$, $i^{\prime}$, $z^{\prime}$
and $J$) as constraints in the SED fitting, which is later discussed in Section
\ref{sec:sed}.
The NUV photometry are based on deep (46 ks) {\it Galaxy Evolution Explorer}
\citep[\GALEX,][]{mar05,mor07} imaging of the SXDS fields (PI: S. Salim, GI6-005).
At $z=0.8$, the NUV band ($\lambda_C \approx 2300$ \AA) samples the rest-frame
far-UV. In addition to the dedicated ``tile'' (1.2-deg circular \GALEX\ pointing)
from this program, our NEWFIRM observations in the SXDS also partially overlap three
shallower archival \GALEX\ tiles with exposure times of 26--30 ks (Figure \ref{fig:1}).
The mean 5$\sigma$ depth for the combined NUV imaging in the NEWFIRM fields is
25.3 mag (exposure time $\sim$80 ks), which is $\sim$1 mag shallower than the deepest
\GALEX\ NUV imaging \citep[the Extended Groth Strip (EGS) field;][]{sal09}, with
integration time of $\approx$260 ks.

Because sources are unresolved in \GALEX\ at $z\approx0.8$ (the \GALEX\ FWHM is
$\sim$5\arcsec), we use PSF source extraction and photometry based on $u$-band priors,
which improves upon the NUV photometry used in \citet{mom13}. The $u$-band photometry
is based on a (1 deg)$^2$ pointing with Canada-France-Hawaii Telescope (CFHT) Megacam
\citep{bou03} with a 3$\sigma$ depth in 2\arcsec\ aperture of 27.0 mag (Foucaud, S.,
private communication)\footnote{Based on publicly available CFHT/Megacam data:
  \url{http://www1.cadc-ccda.hia-iha.nrc-cnrc.gc.ca/en/}.}.
The prior-based photometry is performed using EMphot software 
\citep[version 2.0;][]{vib09}. Priors were limited to $u\leq25$ mag in order to match
the \GALEX\ depth.  We insert artificial sources to check that the fluxes can be
recovered without any systematic errors, and find that not limiting the priors to
$u\leq25$ mag results in an underestimate of the NUV fluxes.  Source extraction was
performed separately in all four tiles (within 0.55 deg radius), and the results were
averaged using photometric errors as weights. The resultant photometry represents the
total galaxy light. NUV (rest-frame FUV) photometry was extracted in areas where
$u$ photometry was available, covering 75\% of NEWFIRM fields (see Figure \ref{fig:1}). 

The optical photometry was obtained with Suprime-Cam \citep{miy02} as part of the
Subaru Telescope Observatory Projects \citep{fur08}. The SXDS fields observed by
NEWFIRM roughly correspond with the Suprime-Cam south, west, and north pointings
(SXDS-S, SXDS-W, and SXDS-N; Figure \ref{fig:1}). Subaru imaging is publicly available
in five broadband filters to 3$\sigma$ depths of $B=28.4$, $V=27.8$, $R_{\rm C}=27.7$,
$i^{\prime} = 27.7$, and $z^{\prime} = 26.6$ mag. The Suprime-Cam data reduction and
photometry are further described in \cite{fur08}.

\section{Stellar Mass, SFR, and Gas-Phase Metallicity Measurements}
\label{sec:prop}
We now turn our attention to calculating physical properties for the galaxies in the
sample from the data presented in Section \ref{sec:data}.

\subsection{Stellar Masses}
\label{sec:sed}

Following the methodology of \citet{sal07}, we derive stellar masses by fitting SEDs to
the eight-band\footnote{Up to 8 bands are available. Coverage in the $u$-band is
  available for \Nfiveagn\ of \Nnoagn\ galaxies, or \percentfiveagn\% of our sample
  (see Figure~\ref{fig:1}).}
photometry that span rest-frame UV and optical wavelengths, their photometric
uncertainties, and spectroscopic redshift.
Stellar masses have already been computed using SED fitting for this same sample for the
nebular reddening analysis of \citet{mom13}.  The main improvement in the calculation
performed here is the inclusion of $u$-band photometry, which in combination with the
NUV photometry, provides more direct constraints on internal dust attenuation for
galaxies at $z\sim0.8$ via the rest-frame UV color.  We give a brief summary of the
modeling used, and refer readers to \citet{sal07,sal09} for further details. 

We use total magnitudes determined within Kron apertures
\citep[MAG\_AUTO from SExtractor;][]{ber96}, except in the NUV band, where PSF extracted
magnitudes are used as described above. The SEDs are fit with a library of 45,000
\cite{bru03} stellar population synthesis models.\footnote{Derived using MAGPHYS
  package available at \url{http://www.iap.fr/magphys/}.}
The model libraries are built with a wide range of SFHs and metallicities, as described
in \citet{sal07}, and updated in \citet{dac08}.
Only models with formation ages lower than 6.8 Gyr, corresponding to the age of the
universe at $z=0.8$, are allowed. Each model is attenuated according to the prescription
of \citet{cha00}, with randomly sampled values of both the total optical depth and the
fraction of the total optical depth due to attenuation by the ambient ISM. The dust
attenuation in the SED fitting is mainly constrained by the UV slope, which gets steeper
with increasing attenuation \citep{cal94}. However, differences in the SFHs can produce
significant scatter between the UV slope and dust attenuation \citep[e.g.,][]{kon04}.
This is generally overcome in our modeling because the near-IR and optical data help to
constrain the age. Intergalactic reddening is included via the prescription of
\citet{mad95}, and a \citetalias{cha03} IMF is assumed. The spectroscopic redshift provides
the luminosity distance, which allows the apparent model quantities to be scaled to
absolute values. We use 0.025 mag calibration errors in all bands, including the NUV
photometry, yielding approximately unit Gaussian residuals with respect to model
photometry. An offset of +0.12 mag in $R_{\rm C}$-band photometry is applied in order to
correct the rest-frame 3700 \AA\ discrepancy with the stellar synthesis models 
\citep{sal09}.

For each galaxy the observed fluxes are compared to those in the model library, and the
goodness of fit ($\chi^2$) determines the probability weight of a given model. The average
of the probability distribution of each fitted parameter is the nominal estimate of that
parameter and its width is used to estimate the errors and confidence intervals. The
majority of galaxies in the sample are well fit and the median $\chi^2$ per degree of
freedom of the best-fitting SED models is close to unity.  In the target redshift range
($0.77<z<0.83$), \Nbadthreeagn\ objects are excluded from the sample due to poor fits
(i.e., if $\chi^2_{\nu}$ of the best-fitting model is $>$10), leaving \Nthreeagn. Another
\Nbadfiveagn\ non-AGN objects lack $u$-band (rest-frame FUV) photometry.

The stellar masses and their uncertainties are given in Table \ref{table:msfr_agn}.
The uncertainties include errors from input photometry and parameter degeneracy (e.g.,
with respect to SFH and dust). Additional systematic uncertainties may arise from the
models themselves and the choice of IMF \citep[e.g.,][]{mara05,con09,tay11}.

\begin{figure}
  \epsscale{1.1}
  \plotone{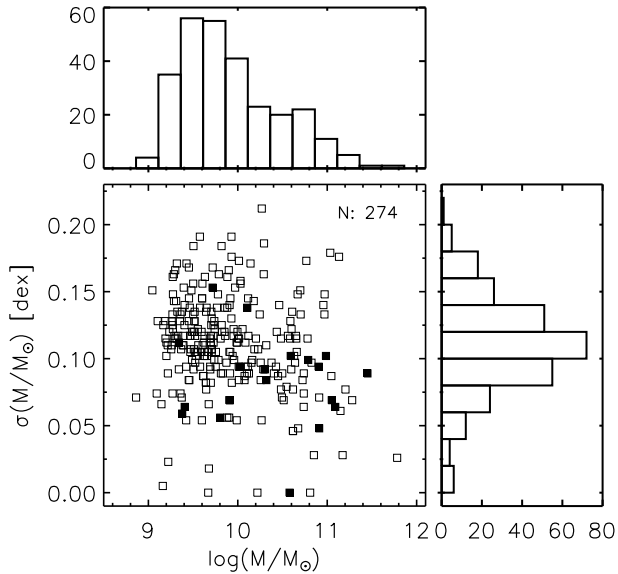}
  \caption{Top: Distribution of stellar masses. Center: Measurement error in stellar
    mass vs. stellar mass (filled points mark AGN). Right: Distribution of stellar
    mass uncertainties.}
  \label{fig:2}
\end{figure}

The distribution of derived stellar masses and their uncertainties are shown in Figure
\ref{fig:2}. AGN are included in the stellar mass sample, but \Nbadthreenew\ sources
(2 are AGN) have poor SED fits and are thus excluded from the original sample size of
\Nspec, leaving \Nthreenew\ sources. The mean (median) of the sample is $10^{9.9}$
($10^{9.8}$) \Msun, with an average uncertainty of $\sigma=0.11$ dex. There are a few
galaxies with masses as low as $10^{8.9}$ \Msun, and as high as $10^{11.8}$ \Msun.

\subsection{Star Formation Rates}
\label{sec:SFR_comp}
The SFRs used in our analysis are measured using two independent methods that are
sensitive to different timescales of star formation.  First, as with stellar masses,
we use the SED fitting to provide a measurement of the recent SFR.  The rest-frame
FUV continuum provides the main constraint on the SFR in the SED modeling, since it
primarily originates from the photospheres of O- through late B-type stars ($M\gtrsim3$
\Msun), and measures star formation averaged over a timescale of $\sim$100 Myr
\citep[e.g.,][]{ken98,lee11}.  The SED modeling also provides constraints on the
attenuation by dust internal to the galaxy, and thus yields dust-corrected SFRs.

\begin{figure}
  \epsscale{1.1}
  \plotone{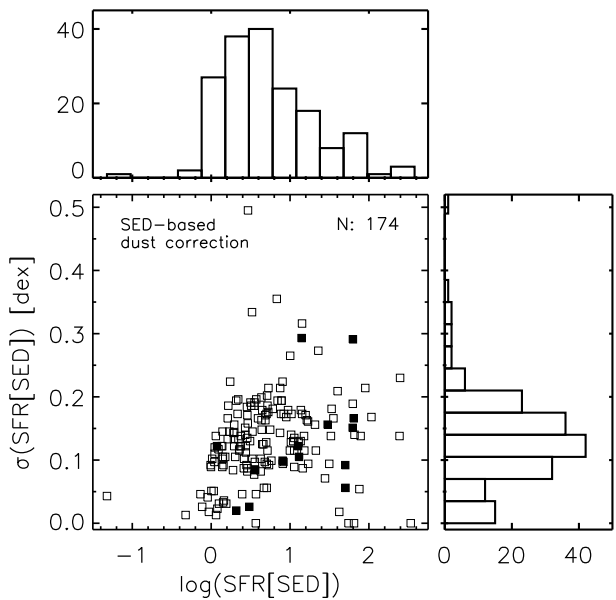}
  \caption{Top: Distribution of dust-corrected UV-based (i.e., from SED fitting) SFRs.
    Center: Measurement error in SFR vs. SFR (filled points mark AGN). Right:
    Distribution of SFR uncertainties.}
  \label{fig:3}
\end{figure}

The distribution of SED-modeled SFRs and their uncertainties are shown in Figure
\ref{fig:3}. As with Figure \ref{fig:2}, the \Nagn\ AGN are included in the sample.
However, \Nbadfivenew\ sources lacked $u$-band observations, leaving a sample size of
\Nfivenew. The SED-modeled dust-corrected SFRs have a mean (median) of $10^{0.72}$
($10^{0.64}$)  \Msun\ yr$^{-1}$, ranging from $10^{-1.3}$ to $10^{2.5}$  \Msun\ yr$^{-1}$
with an average uncertainty of $\sigma=0.13$ dex.

Second, we compute SFRs using the H$\alpha$ luminosities derived from the NEWFIRM
narrowband photometry.  H$\alpha$ nebular emission directly arises from the
recombination of \textsc{H ii} gas ionized by the most massive O- and early B-type
stars ($M\gtrsim10$ \Msun), and therefore traces star formation over the lifetimes of
these stars, which is on the order of a few million years \citep[e.g.,][]{ken98}.
The H$\alpha$ SFRs are computed as follows.  H$\alpha$+\NII\ fluxes are first calculated
from the $J$ and NB118 photometry, as described in \cite{ly11a}. The fluxes are
converted to luminosities using the spectroscopic redshifts. Corrections must then be
applied for: (1) the contribution of the
\NII\,$\lambda\lambda$6548,6583\footnote{Hereafter, ``\NII'' refers to the sum of the
two nitrogen emission lines, unless otherwise indicated.} lines, and (2) attenuation
by dust internal to the galaxy.

\begin{figure*}
  \epsscale{1.1}
  \plottwo{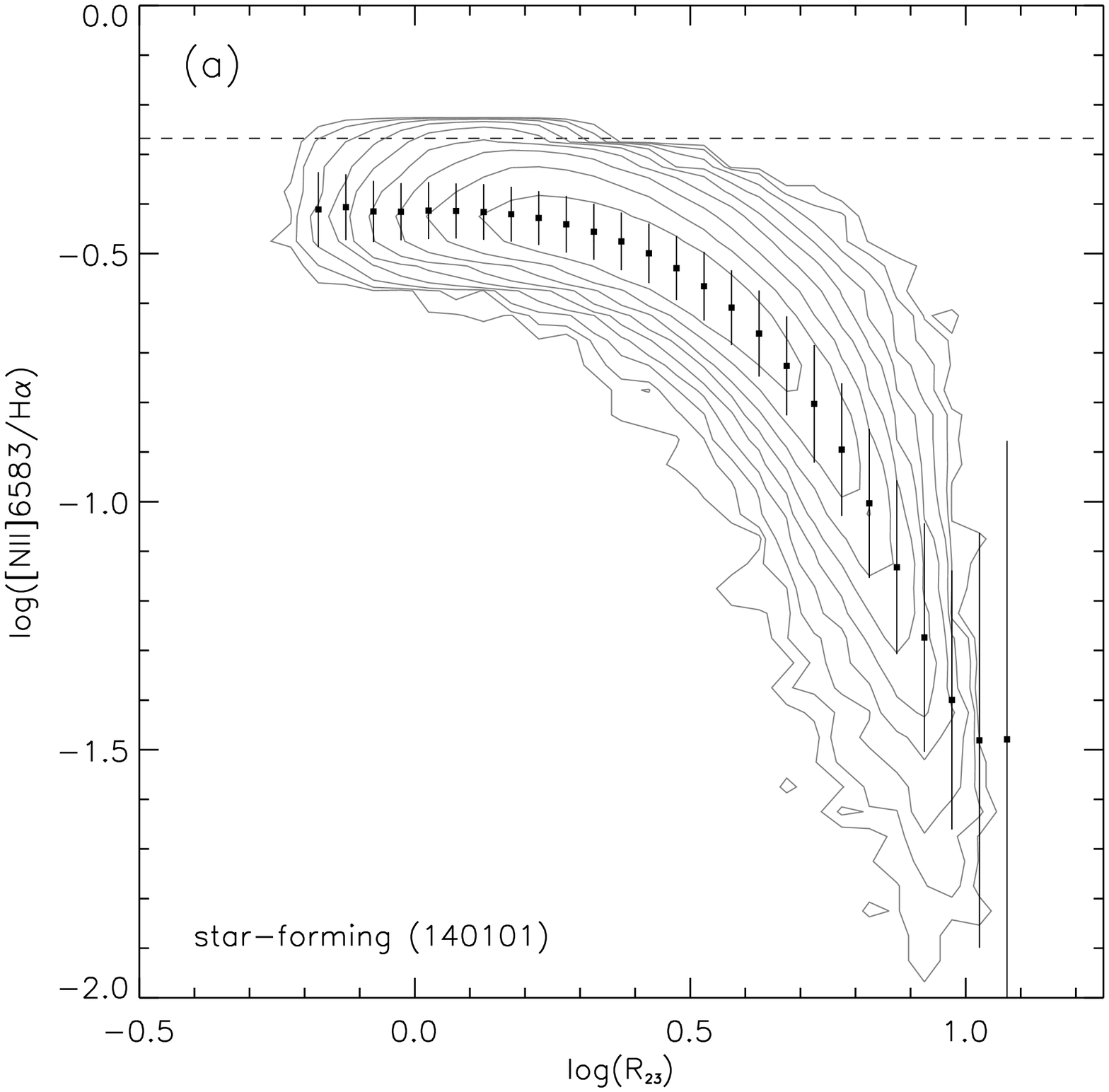}{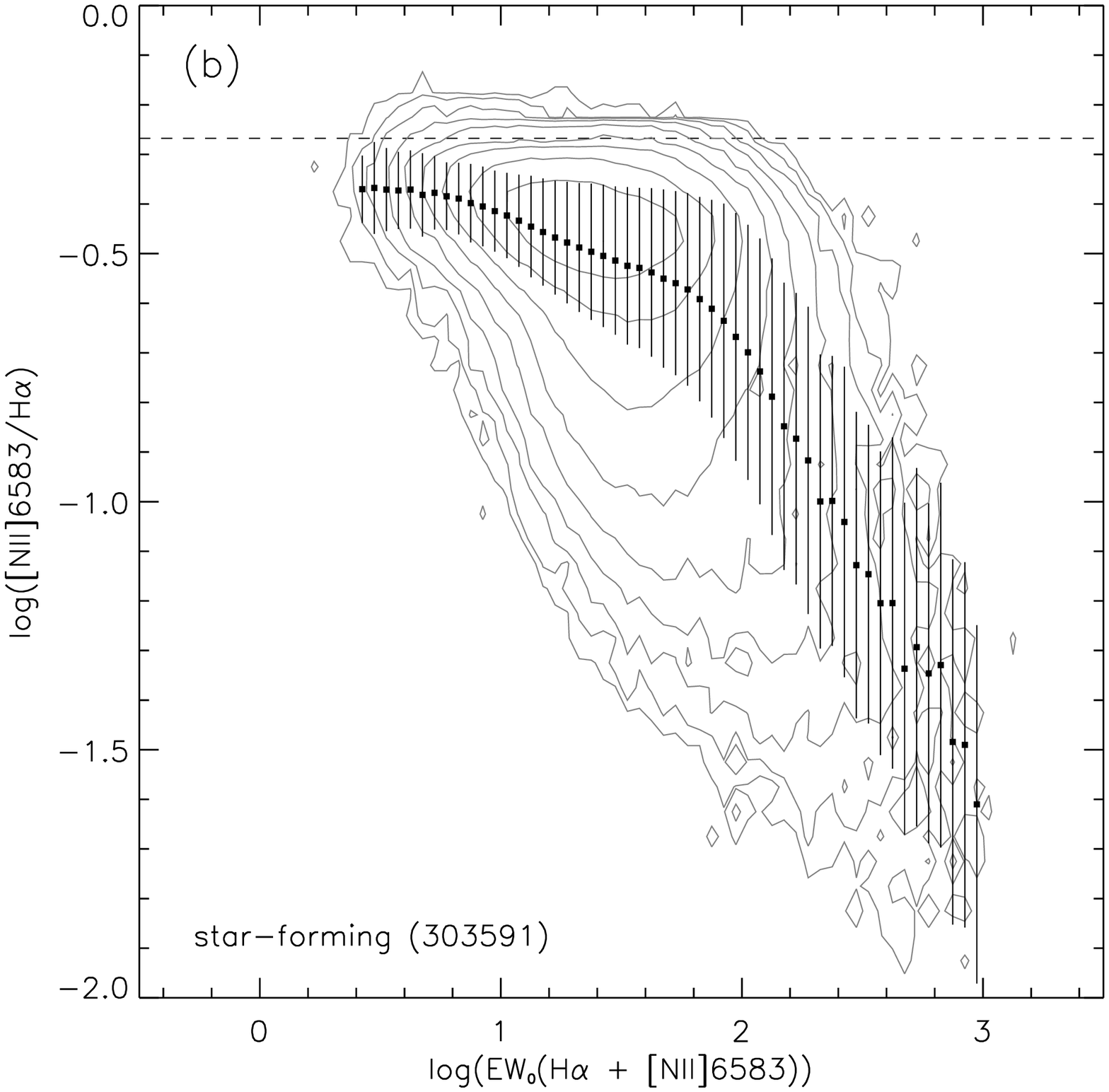}
  \caption{\NII\,$\lambda$6583/\Ha\ as a function of (a) $\log(R_{23})$ and
    (b) $\log({\rm EW_o}({\rm H}\alpha + [\textsc{N ii}]\,\lambda$6583)) from
    the SDSS DR7 sample.
    Contour levels are on a logarithmic scale. The average and 1-$\sigma$
    dispersion are shown by the filled squares and errorbars. The dashed
    lines refer to the maximum \NII\,$\lambda$6583/\Ha\ ratio of 0.54
    for star-forming galaxies \citep{ken08}. These correlations are used
    to remove contamination of \NII\ in our NB118 bandpass to yield
    \Ha-based SFRs.}
  \label{fig:4} 
\end{figure*}

\noindent{\it \textsc{N ii} contamination.} The NB118 bandpass is wide enough to include
flux from the \NII\ emission lines for \Ha\ excess emitters. To correct for this, we
estimate the \NII/\Ha\ ratio with the $R_{23}$ flux ratio \citep{pag79},
\begin{equation}
    R_{23} \equiv \frac{\OII\,\lambda\lambda3726,3729+\OIII\,\lambda\lambda4959,5007}
    {{\rm H}\beta}.
\end{equation}
We follow this empirical approach since both the \NII\,$\lambda$6583/\Ha\ and $R_{23}$
are often used to determine metallicity \citep[e.g.,][see Section \ref{sec:metal} for
further discussion]{pet04,kew08}, so a tight correlation is expected. To calibrate
this method, we begin with the largest spectroscopic sample, the SDSS MPA-JHU Data
Release 7 \citep[DR7;][]{aba09} catalog.\footnote{\url{http://www.mpa-garching.mpg.de/SDSS/DR7/}.}
We require that \OIII\,$\lambda$5007, H$\beta$, H$\alpha$, and both the \OII\ lines to
be detected at a minimum of 3$\sigma$. This restriction limits the DR7 sample to 165,622
galaxies. We do not apply a restriction on \NII, as the line is intrinsically weak, so
a required detection will bias the correction against metal-poor galaxies.
We then use the \cite{bal81} ``BPT'' diagnostic diagram to exclude AGN. Here we adopt the
\cite{kau03} selection for star-forming galaxies, which limits the sample further to
140,101 galaxies. We illustrate the \NII\,$\lambda$6583/\Ha\ and $R_{23}$ ratios for
these galaxies in Figure \ref{fig:4}(a).
It can be seen that the two ratios are well-correlated with \NII\,$\lambda$6583/\Ha\
reaching a maximum of $\approx$0.4.
To correct for \NII/\Ha, we use the mean values, which are shown as solid filled squares
in Figure \ref{fig:4}(a). We also factor in the dispersion of this correlation, which is
typically $\sigma\lesssim0.1$ dex.
Since the \NII\,$\lambda$6583 is the stronger of the two \NII\ lines, we also assume
$\lambda$6583/$\lambda$6548 = 3. For our sample, the \NII\ correction is between
$\log(\NII/{\rm H}\alpha)$ = --1.36 and --0.29, with an average of --0.55.

We note that for a subset of our galaxies ($N$ = 168), the emission lines used for
  computing $R_{23}$ are not well measured ($<$3$\sigma$). To correct these galaxies
for their \NII\ contamination, we follow previous efforts that use the \Ha+\NII\
equivalent width (EW). This method was first implemented by \cite{vil08}.
One problem with the previous calibration was the inclusion of AGN and LINERs
\citep{hec80}, which significantly biased the \NII\ correction. Here, we therefore
reproduce the relation of \cite{vil08} with only star-forming galaxies.
We emphasize that the AGN contribution is low in our sample, and we have utilized various
empirical methods to identify and remove AGN from our sample (see Section \ref{sec:imacs}).
The EW correlation of \NII/\Ha\ is illustrated in Figure \ref{fig:4}(b).

We find that both the $R_{23}$- and EW-based methods yield fairly consistent \NII\
corrections for the sample with $R_{23}$ emission lines (\OII, \OIII, and H$\beta$)
detected at $\geq$3$\sigma$; the EW-based \NII\ corrections are higher than the $R_{23}$
\NII\ corrections by $\sim$0.05 dex, with a dispersion of $\sim$0.03 dex.
We also find that the EW approach suffers from greater dispersion. This is not a surprise
since a tight correlation is not expected between the specific SFR (SFR per unit stellar
mass; SFR/\Mstar) of galaxies, as measured from the \Ha\ EW, and their metallicity, as
measured from \NII\,$\lambda$6583/\Ha.
Since previous studies \citep[e.g.,][]{sob09} used the \cite{vil08} calibration, we note
that their \Ha\ measurements are underestimated due to a systematically larger correction
for \NII.

We note that adopting local measurements to correct for \NII\ contribution in the NB
filter has its limitations, particularly since a few studies of strongly star-forming
galaxies have seen nitrogen abundance enhancements relative to oxygen
\citep[][]{amo10,mas14}. This result is not too surprising since nitrogen has a secondary
production source. Given this recent evidence, it is therefore likely that we are
underestimating the \NII\ contribution, and thus overestimating the \Ha\ flux. For the
purpose of our analyses, we defer on this issue, as we plan to revisit it in future work.

\noindent{\it Dust attenuation.} The H$\alpha$ emission-line luminosities are corrected
for dust reddening in two ways: using both the Balmer decrement (H$\gamma$/H$\beta$)
from spectroscopy and the estimate of nebular attenuation from SED fitting. Since the
H$\gamma$ line is intrinsically weak for much of our sample, the majority of the
following analysis is based on dereddening based on SED-derived attenuation.

\begin{figure}
  \epsscale{1.1}
  \plotone{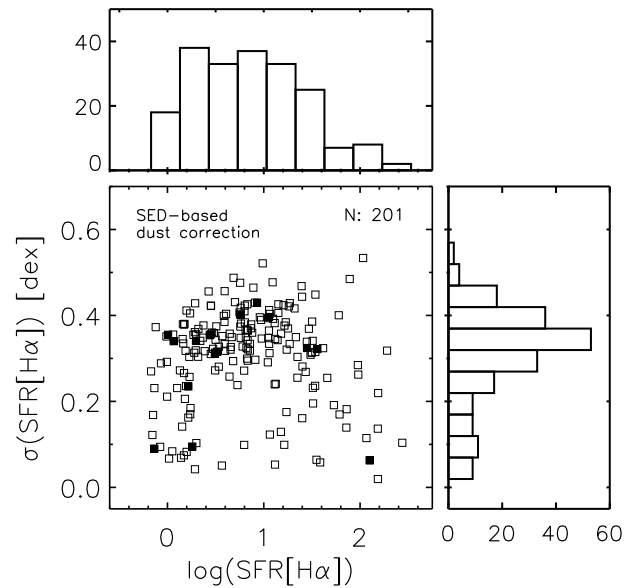}
  \caption{Top: Distribution of dust-corrected \Ha-based SFRs. Center: Measurement error
    in SFR vs. SFR (filled points mark AGN). Right: Distribution of SFR uncertainties.}
  \label{fig:5}
\end{figure}

Finally, we use the prescription of \cite{ken98} to derive SFRs from \Ha\ luminosities.
We divide by a factor of 1.8 to convert the SFRs from a \cite{sal55} to a
\citetalias{cha03} IMF. The distribution of dust-corrected \Ha-based SFR
\citep[corrected using SED results and the extinction formalism of][]{cha00} is shown
in Figure \ref{fig:5}. AGN are again included in this sample; however, 90 sources
with $<3\sigma$ NB118 excess flux are removed and 8 more sources without SED fits,
leaving a sample size of \Nonefournew. The mean (median) of the sample is
$10^{0.74}$ ($10^{0.67}$)  \Msun\ yr$^{-1}$,
and the average uncertainty is $\sigma=0.33$ dex. There are a few galaxies with SFRs
as low as $10^{-0.4}$ \Msun\ yr$^{-1}$, and as high as $10^{2.4}$ \Msun\ yr$^{-1}$. SFRs
based on the SED-fitting and \Ha\ luminosities are reported in Table \ref{table:msfr_agn}.

\subsection{Gas-Phase Metallicities}
\label{sec:metal}
Various metallicity calibrations have been developed for over two decades, yet the
absolute metallicity scale is still uncertain, as demonstrated by \cite{kew08}. In
general, oxygen abundance is used as a proxy for global gas-phase metallicity and
expressed as a dimensionless quantity, $Z \equiv 12+\log{({\rm O/H})}$. On this
scale, $Z_{\odot}=8.76$ \citep{caf11}.

The ``direct'' method of determining $Z$ is to measure the ratio of the weak \OIII\
$\lambda$4363 line to a lower excitation line, which gives an estimate of the electron
temperature $T_e$ that is inversely related to the gas metallicity. While efforts
have measured direct metal abundances \citep[e.g.,][]{kak07,bro08,hu09,ber12,ly14},
\OIII\ $\lambda$4363 is very difficult to robustly detect.
As a result, strong-line calibrations based on the empirical relationship between
$T_e$-based metallicities and strong-line ratios (e.g., $R_{23}$) have been developed
\citep[e.g.,][]{n06}. Other calibrations use population synthesis and photoionization
models to calculate theoretical strong-line ratios for various input metallicities
\citep[e.g.,][hereafter Z94, M91 and KK04, respectively]{z94, m91, kk04}.
Finally, Bayesian fitting has been used to find the photoionization model that best
explains the observed fluxes of all the most prominent rest-frame optical emission
lines \citepalias{tre04}. 

Metallicities determined from the direct and empirical methods based on $T_e$ have
been shown to be systematically lower than those determined from the theoretical
methods based on photoionization models \citep{kew08}.
While this discrepancy is unresolved, problems with photoionization models, or
temperature gradients/inhomogeneities may cause $T_e$ methods to underestimate true
metallicities \citep{kew08}. Also, recent efforts have suggested that non-Maxwellian
energy distributions in the ISM may be the culprit for many systematic differences
\citep{nic12, nic13,dop13}. Regardless of what is responsible for the discrepancies,
it is clear that a consistent use of a single metallicity calibration is required
to obtain a self-consistent \MZS\ relation and to study its evolution.

For completeness and to aid direct comparisons in future work, we have 
determined metallicity using multiple calibrations.
In Tables \ref{table:z_sed_3sig} and \ref{table:z_bal_3sig}, we present
these metallicities, calculated using the nebular emission lines reported in
Table \ref{table:emline_flux}, with dust attenuation correction derived from
the SED fitting and the Balmer decrement (H$\gamma$/H$\beta$), respectively.
The transformation from observed line ratios to gas metallicity is summarized
in the Appendix of \cite{kew08}.

Note that several calibrations rely on the $R_{23}$ line ratio \citep{pag79}. The
\citetalias{m91} and \citetalias{kk04} calibrations also use the $O_{32}$ line ratio:
$O_{32} \equiv$ \OIII\,$\lambda\lambda$4959,5007/\OII\,$\lambda\lambda$3726,3729,
as an estimate of the ionization state of the gas. 
This measurement helps to resolve the degeneracy between ionization state and
metallicity (see e.g., Figures 3 and 12 of \citetalias{kk04} and \citetalias{m91},
respectively).
For this reason, we choose to use the \citetalias{m91} calibration in analyzing the
\MZ\ relation and comparing our work with previous results (Section~\ref{sec:MZR}).
For instances where a different metallicity calibration was used, we converted
to \citetalias{m91}-based metallicity using the relations defined in Table 3 of
\cite{kew08}.
 
For the \MZS\ relation (Section~\ref{sec:mzs_relation}), previous studies have used
the \citetalias{tre04} calibration rather than \citetalias{m91}
\citepalias[e.g.,][]{lar10,yat12}. To aid in direct comparisons, we analyze the
\MZS\ relation using metallicities scaled with respect to the photoionization models
derived by \citetalias{tre04}; 
however, supplementary results using \citetalias{m91} metallicities are provided. 
When computing \citetalias{tre04}-based metallicities, we use the following 
empirical $R_{23}$--$Z$ relation provided by \citetalias{tre04}:
\begin{equation}
  12+\log{\left({\rm O/H}\right)} = 9.185 - 0.313x -0.264x^2 - 0.321x^3,
\end{equation}
where $x \equiv \log{(R_{23})}$.

\begin{figure}
  \epsscale{1.1}
  \plotone{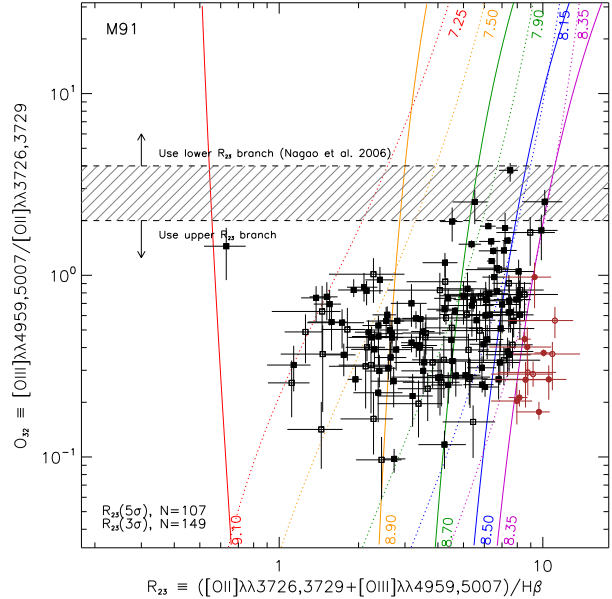}
  \caption{Metallicity-sensitive ($R_{23}$) and ionization parameter-sensitive ($O_{32}$)
    emission-line ratios for the $z=0.8$ New\Ha\ sample. We limit the sample to galaxies
    that are not AGN, as well as, galaxies that have reliable SED fits (i.e., these
    galaxies are later used to construct the \MZ\ relation). Filled points show the
    $R_{23}(5\sigma)$ sample with additional sources from the $R_{23}(3\sigma)$ sample as
    unfilled points. Circles (brown in the color version) indicate galaxies with high
    $R_{23}$ values such that the upper branch metallicity is less than the lower branch
    metallicity. Photoionization models from \citetalias{m91} are overlaid in colors for
    metallicities between 12 + log(O/H) = 7.25 and 12+log(O/H) = 9.1. Solid (dotted) curves
    are for metallicities on the upper (lower) $R_{23}$ branch. Based on the empirical
    relations of \cite{n06}, the dashed horizontal lines distinguish between upper and
    lower $R_{23}$ branch with a region of ambiguity (shaded). Given these $R_{23}$ and
    $O_{32}$ values, we adopt the upper branch.} 
  \label{fig:6}
\end{figure}

The relationship between $R_{23}$ and O/H is double-valued, and so a given value of
$R_{23}$ may correspond to low or high metallicity (``lower branch'' and ``upper branch'',
respectively), and additional line ratios are needed to break the degeneracy.  
One such ratio is \NII\,$\lambda$6583/\Ha\ \citep{kew08}; however, this requires
medium-resolution infrared spectroscopy for our galaxies, which we currently do not
have. An alternative line ratio that can distinguish between the upper and lower
branch is the $O_{32}$ ratio \citep[e.g.,][]{n06,mai08}. We illustrate the
dust-corrected $R_{23}$ and $O_{32}$ ratios in Figure \ref{fig:6}, which demonstrates
that the majority of our sources follow the upper $R_{23}$ branch. There are a few
sources with line ratios such that the choice in upper or lower branch is ambiguous
(shaded region in Figure \ref{fig:6}). Adopting either branch does not
impact the primary results of our \MZ--(SFR) analysis.

\begin{figure}
  \epsscale{1.1}
  \plotone{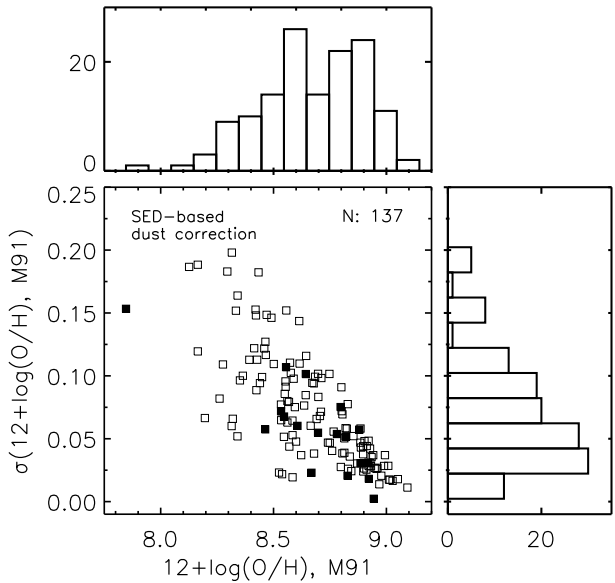}
  \caption{Top: Distribution of \citetalias{m91} upper-branch metallicities. Center:
    Measurement error in \citetalias{m91} upper-branch metallicity vs. \citetalias{m91}
    upper-branch metallicity (filled points mark AGN). Right: Distribution of
    \citetalias{m91} upper-branch metallicities uncertainties. Only galaxies
      with \OII, \OIII, and \Hb\ $>5\sigma$} are shown here.
  \label{fig:7}
\end{figure}

The distribution of M91 upper-branch metallicities and their errors 
\citep[corrected using SED results and the extinction formalism of][]{cha00} are shown
in Figure \ref{fig:7}. AGN are included in this figure, but the sample is
restricted to the \Nfivesignew\ sources where \OII, \OIII, and H$\beta$ are all
detected at $\geq$5$\sigma$. We refer to this sample as ``$R_{23}(5\sigma)$'' for
brevity. This naming convention should not be interpreted as a 5$\sigma$ detection
limit on $R_{23}$, rather on the emission lines for the line ratio.
A larger sample of \Nthreesignew\ galaxies, selected at 3$\sigma$, is also constructed,
hereafter ``$R_{23}(3\sigma)$''.
The mean (median) of the sample is $12+\log({\rm O/H})$ = 8.67 (8.68), and the average
uncertainty is 0.07 dex. There are a few galaxies with metallicities as low as 7.85,
and as high as 9.09.

\section{Results and Analysis}
\label{sec:analysis}
In our analysis, we first compare the two measures of the SFR computed above, those
based upon the H$\alpha$ flux, and those derived from the SED modeling of rest-frame
FUV to $R$-band photometry, as a check on the relative reliability of the methods
(Section \ref{sec:SFR}).  We then compare the stellar mass, SFR, and metallicity
measurements for the New\Ha\ sample to other measurements in the literature, as
studied within the framework of the star-formation sequence (Section \ref{sec:mainseq})
and the \MZ\ relation (Section \ref{sec:MZR}). Finally, we combine these two relations
to investigate the \MZS\ relation at $z\approx0.8$ (Section \ref{sec:mzs_relation}).

In each analysis, we use various combinations of sample restrictions. For convenience,
the sample cuts used and the subsample sizes are compiled in Table~\ref{table:subsample}.

\subsection{Comparison of SFR Tracers}
\label{sec:SFR}
In order to compare SFR measurements from different tracers, we do not consider AGN. From
the remaining  sample of \Nnoagn\ galaxies, only those with an NB118 excess line flux
$\geq3\sigma$ and good SED fits ($\chi_{\nu}^2<10$) with $u$-band photometry are included,
leaving a sample size of \NSFRcompare.

\begin{figure}
  \epsscale{1.1}
  \plotone{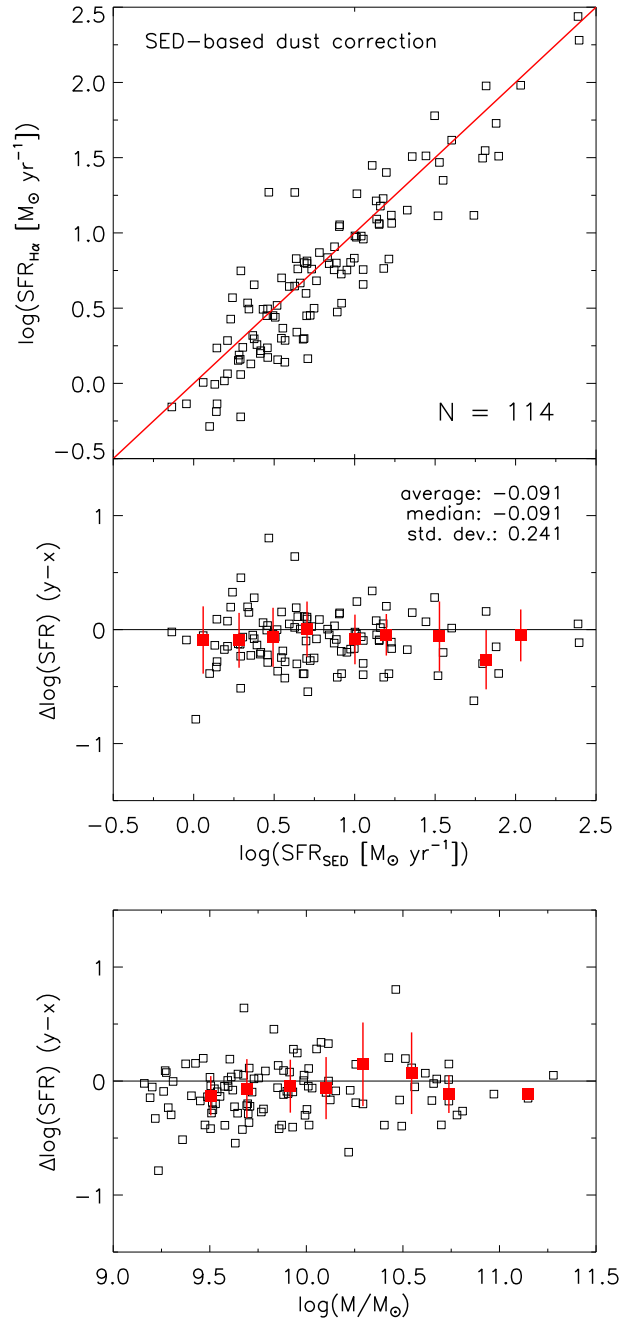}
  \caption{Comparison of SED-based and \Ha-based SFRs, both corrected for extinction using
    SED results and the extinction law of \citet{cha00}. Solid lines (red in
    online version) indicate one-to-one. Filled points (red in online version) in the
    middle and bottom panels indicate median values within SED-based SFR and mass bins,
    respectively, while the errorbars represent the 1$\sigma$ scatter in the residuals.}
  \label{fig:8}
\end{figure}

We illustrate the SED- and \Ha-based SFRs in Figure \ref{fig:8}. Since the SED modeling
constrains the dust attenuation, we also correct the \Ha\ measurements with the SED-based
dust extinction estimates for each of our galaxies, $A$(\Ha) = 0.96$\tau_V$.
We find that the SED-based SFRs are higher than the \Ha-based SFRs by $\sim$0.09 dex
in the median, well within the scatter of 0.24 dex. In addition, we find that there is no
mass dependence of the residual, and that these two measurements even agree at high SFRs
($\gtrsim$100 \Msun\ yr$^{-1}$). We note that similar results on the SFR comparison are
found using Balmer decrements (H$\gamma$/H$\beta$, obtained from spectroscopy;
Section~\ref{sec:imacs}) for dust attenuation. However, only 27 galaxies have
deep enough spectra (H$\gamma$ SN $\geq10$) to yield individual decrements that are
reliable at $\Delta(E_{B-V})=0.2$ mag.

These results are roughly consistent with other comparisons of SFR tracers at similar
or lower redshifts. For example, \cite{vil11} compared H$\alpha$- and FUV-based SFRs at
$z\sim0.84$. While \citet{vil11} found that observed H$\alpha$-based SFRs were
systematically higher than FUV-based SFRs, they also found that correcting for dust
caused the two SFR tracers to agree at the level of 0.05 dex with a dispersion of
$\approx$0.2--0.25 dex.
Also, \cite{ly12} have compared SED-based SFRs against \Ha\ SFRs in $z=0.4$--0.5
H$\alpha$-selected galaxies, and also find good agreement with low dispersion
($\sim$0.2 dex) with corrections for dust attenuation based on estimates from SED fits.

Since the H$\alpha$-based SFRs are more robust to the effects of dust attenuation
compared to the SED-based SFRs, we hereafter use \Ha\ SFRs, corrected for dust
attenuation determined from SED fitting.

\subsection{The Star-Formation Sequence}
\label{sec:mainseq}
The relation between SFR and \Mstar, commonly referred to as the ``star-forming sequence''
\citep{sal07} or the ``main sequence of star-forming galaxies'' \citep{noe07}, has been
well studied at low \citep[e.g.,][]{bri04, sal07} and intermediate
\citep[e.g.,][]{noe07, vil11, whi12} redshifts, with general agreement among different
literature results. It manifests as a relatively tight ($\sigma\sim0.3$ dex) relationship
usually parameterized as SFR $\propto M_{\star}^{\beta}$, although some recent
  works have suggested that the \MSFR\ relation is not a simple power law at high
  redshifts \citep{whi14}.
The evolution of the form of the \MSFR\ relation over cosmic time can provide constraints
on the characteristic star formation history of galaxies and the significance of
episodic bursts of activity in building up the stellar mass \citep[e.g.,][]{noe07}.
In this section, we compare our NewH$\alpha$ results with other studies at $z\sim0.8$.

\begin{figure}
  \epsscale{1.1}
  \plotone{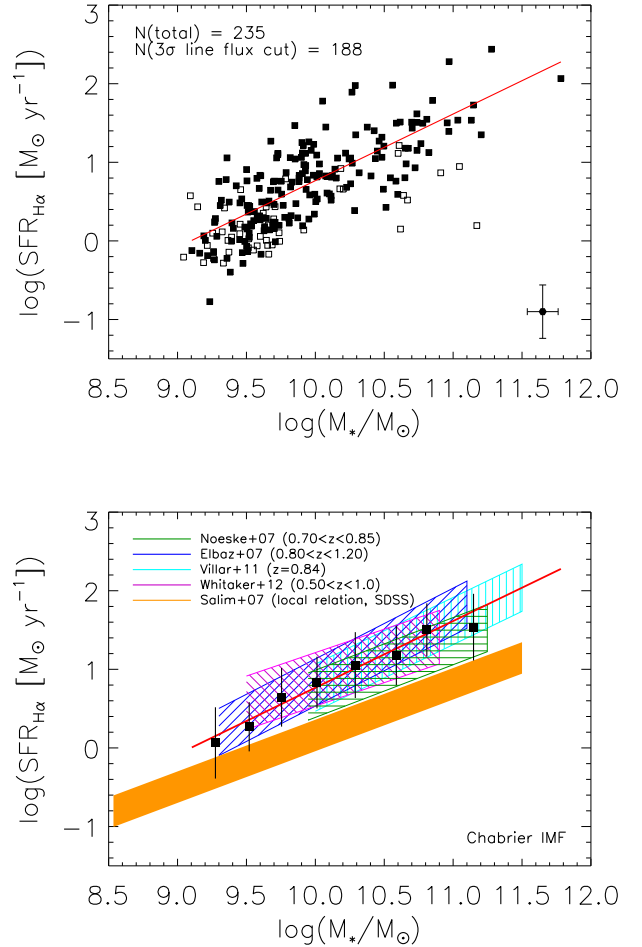}
  \caption{Top: Comparison of the NewH$\alpha$ \Mstar--SFR data (black squares) with a
    linear fit (solid line, red in the online version). Filled (open) squares mark
    sources for which the NB118 excess flux is $>3\sigma$ ($<$3$\sigma$); the
    linear fit uses the $>$3$\sigma$ points. Bottom: Comparison of the NewH$\alpha$
    \Mstar--SFR linear relation with literature relations. Black points denote medians
    of NewH$\alpha$ data binned by mass, with errorbars representing standard
    deviations. Note that the linear fit covers the entire mass range, while the black
    points only mark mass bins with more than one data point.}
  \label{fig:9}
\end{figure}

The top panel in Figure \ref{fig:9} plots the NewH$\alpha$ \Mstar--SFR data and a
least-squares linear fit to our 3$\sigma$ sample (i.e., NB118 excess flux $\geq3\sigma$).
We find the resulting best-fit power law to be:
\begin{equation}
\log{\left(\frac{\SFRHa}{M_{\odot}~{\rm yr}^{-1}}\right)} =
(0.75\pm0.07)\log{\left(\frac{M_{\star}}{M_{\odot}}\right)}-(6.73\pm0.67).
\label{eqn:mainseq}
\end{equation}

The bottom panel compares this linear fit (red line with average measured scatter of
0.47 dex) against literature star-formation sequences and estimated intrinsic scatters
determined by \citet{noe07} and \citet{elb07} \citep[compiled by][]{dut10}, as well as
\citet{vil11} and \citet{whi12}. The medians of the NewH$\alpha$ data binned by mass
(black points) are provided in Table~\ref{table:binned}. When necessary, stellar masses
and SFRs have been converted to a \citetalias{cha03} IMF from a \cite{sal55} IMF by
dividing by 1.8 \citep[e.g.][]{gon10}. The local star-formation sequence
\citep[solid orange band]{sal07,sal12} has also been plotted for comparison.

Despite small variations, the NewH$\alpha$ \Mstar--SFR relation and literature relations
at similar redshift are all systematically higher than the local relation \citep{sal07}.
This is consistent with studies that generally find that galaxies at higher redshift
tend to have higher SFRs at fixed stellar mass compared to galaxies in the local universe.

As noted in Section \ref{sec:sed}, the NewH$\alpha$ survey covers a stellar mass range
of $10^{8.9}$ to $10^{11.8}$ \Msun\ with an average mass of $10^{9.9}$ \Msun. At similar
redshifts, \citet{noe07} investigated a limited stellar mass range of
$\sim10^{10.0}$--$10^{11.3}$ \Msun, while \citet{elb07} covered a mass range of
$\sim10^{9.3}$--$10^{11.1}$ \Msun\ and \citet{whi12} covered a mass range of
$\sim10^{9.5}$--$10^{11.0}$ \Msun. It is notable that despite varying mass ranges, sample
selection methods and SFR determinations, \citet{noe07}, \citet{elb07}, and \citet{whi12}
each find best-fit power-law relationships that are all fairly consistent with each
other. The NewH$\alpha$ fit also agrees well with these relations, with some minor
variations.

The \citet{noe07} relation (green band horizontally cross-hatched) is
$\log{(\rm SFR/(M_{\sun}~{\rm yr}^{-1}))}=(0.67\pm0.08)\log{(M_{\star}/M_{\sun})}-(5.96\pm0.78)$
\citep[using the relation compiled by][]{dut10}. It therefore has a shallower slope than
the NewH$\alpha$ relation. However, the NewH$\alpha$ fit lies within the scatter
of the \citet{noe07} relation ($\sim$0.3 dex). The slope of the \citet{elb07} relation
(blue band cross-hatched diagonally upwards right) is $0.90$, rising more steeply
than our fit, but its intercept is much lower \citep[--8.17, again using the relation
compiled by][]{dut10}. This could simply be due to the fact that \citet{elb07} study
galaxies at a slightly higher median redshift and wider redshift range ($0.80<z<1.20$).
Despite this offset, the NewH$\alpha$ fit is within the scatter of the \citet{elb07}
relation ($\sim0.3$ dex). On the other hand, the \citet{whi12} relation
(pink band cross-hatched diagonally downwards right) has a shallower slope, with a
functional form of $\log{(\rm SFR/(M_{\sun}~{\rm yr}^{-1}))}=0.6\log{(M_{\star}/M_{\sun})}-5.09$
at $z\sim0.8$ (uncertainties not given). This is more likely due to incompleteness at
masses below $10^{9.8}$ \Msun, as shown in Figure 1 of \citet{whi12}. Above this
mass-completeness limit, the NewH$\alpha$ relation is well within the scatter derived by
\citet{whi12} ($\sim0.34$ dex).

\citet{vil11} use a narrowband selection similar to that used in the NewH$\alpha$ survey
and study 153 H$\alpha$ emitters at $z\sim0.84$ in the EGS and GOODS-North fields. This
survey covered a mass range of $10^{10.0}$--$10^{11.5}$ \Msun. \citet{vil11} do not
provide a functional form for their \Mstar-SFR relation; however, when we calculated
our own least-squares linear fit for the \citet{vil11} data points (light-blue
band cross-hatched vertically), we found a fit of
$\log{(\rm SFR/(M_{\sun}~{\rm yr}^{-1}))}=(0.51\pm0.15)\log{(M_{\star}/M_{\sun})}-(4.11\pm1.61)$.
Although both the \citet{vil11} slope and intercept are systematically lower than the
NewH$\alpha$ slope and intercept, the NewH$\alpha$ relation is consistent with the
\citet{vil11} relation at overlapping mass bins (i.e., above $10^{10}$ \Msun).

Since our sample is selected by \Ha\ emission, we expected our galaxies to be biased
toward higher SFRs, which would result in a \Mstar--SFR relation systematically higher
than those found using mass- and luminosity-limited surveys. However, that is not the
case, as demonstrated in Figure~\ref{fig:9}. This is because our \Ha\ survey is
reasonably deep with a 50\% completeness limit that corresponds to an observed \Ha\
SFR of 0.4 \Msun\ yr$^{-1}$ \citep{ly11a}. Our greater observational limitation is on
the amount of excess flux (i.e., the \Ha\ EW) that we can measure, which corresponds
to the specific SFR. Previous Monte Carlo simulations suggest that the low-EW
population that we are missing amounts to $\approx$20\% at high masses and increases
to 50\% near our sensitivity limits \citep{ly11a}. These selection limitations, however,
do not appear to bias our sample any more than mass- and luminosity-limited surveys. We
note that \cite{hen13b}, who used a sample of emission-line galaxies selected from grism
spectroscopy, have also found good agreement with other \Mstar--SFR studies.

We also note that despite the consistency between the New\Ha\ data with
literature \MSFR\ relations, the upper panel of Figure~\ref{fig:9} shows that a line
does not perfectly fit the New\Ha\ \MSFR\ data. This may be evidence that the \MSFR\
relation is not a simple power law at higher redshifts, as recently suggested by
\citet{whi14}. However, the observed curvature in the \MSFR\ relation may be the
result of selection bias (i.e., missing more dust-obscured galaxies) or because the
dust attenuation correction is underestimated.

\subsection{The Mass--Metallicity Relation}
\label{sec:MZR}
Several previous studies have examined the \MZ\ relation at $z\sim0.8$. However, these
studies used samples with different ranges of stellar mass and metallicity, as well
as, different treatments of systematic effects like dust extinction
\citep[for some analysis, see][]{mou11,zah11}. We therefore aim to supplement and
compare previous results with our large spectroscopically-selected NewH$\alpha$ sample.

\begin{figure}
  \epsscale{1.1}
  \plotone{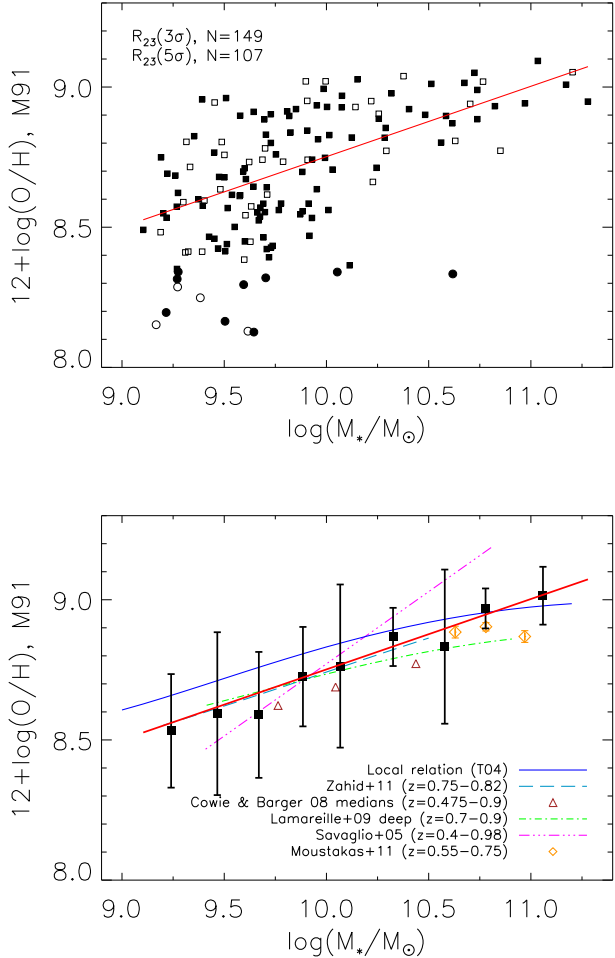}
  \caption{Top: Comparison of the NewH$\alpha$ \MZ\ data with a linear fit
    (solid line, red in online version). Filled points mark sources for which
    $R_{23}$ lines (\OII, \OII, and \Hb) are at least detected at 5$\sigma$ with
    additional sources from the 3$\sigma$ sample (unfilled). Circles indicate galaxies with high
    $R_{23}$ values such that the upper branch metallicity is less than the lower branch
    metallicity. Bottom: Comparison of the NewH$\alpha$ \MZ\ linear fit with
    literature relations. All metallicities have been converted to the \citetalias{m91}
    calibration using the relations of \citet{kew08}; for the local relation, we
    transformed from \citetalias{tre04}. Filled squares denote means of NewH$\alpha$
    data binned by mass, with errorbars representing standard deviations.}
  \label{fig:10}
\end{figure}

As before, we fit the NewH$\alpha$ data with a linear relation using least-squares. 
We plot both $R_{23}(3\sigma)$ and $R_{23}(5\sigma)$ metallicity detections,
but for the linear fit, we use the $R_{23}(5\sigma)$ sample. Using the 5$\sigma$ cut did
not significantly bias the sample, as the top panel of Figure \ref{fig:10} shows.
We also remove 9 sources for which the calculated upper-branch metallicity is lower
than the lower-branch metallicity, leaving a 5$\sigma$ sample size of
\Nmzfivesig\footnote{These 9 galaxies have $R_{23}$ values that are higher than the 
  limit of many of these metallicity calibrations (see Section \ref{sec:metal} and
  Figure \ref{fig:6}). \citet{zah11} removed such galaxies as well, believing them to
  be AGN.}. 
Using the \citetalias{m91} metallicity calibration, we find the resulting linear
relation to be \begin{equation}
12+\log{({\rm O/H})} = (0.25\pm0.03)\log{\left(\frac{M_{\star}}{M_{\odot}}\right)}+
(6.23\pm0.33),
\label{eqn:MZ}
\end{equation}
with an intrinsic scatter of 0.16 dex.

The top plot in Figure \ref{fig:10} shows this linear relation, while the bottom plot
compares this linear fit to other \MZ\ relation determinations at similar redshifts.
The means of the NewH$\alpha$ data binned by mass are also plotted, and their values
are provided in Table~\ref{table:binned}. All stellar masses are converted to be
consistent with a \citetalias{cha03} IMF, and all metallicities are made to be consistent
with the \citetalias{m91} upper-branch calibration using the conversions in \cite{kew08}.

We find that the NewH$\alpha$ \MZ\ relation is generally consistent with literature
relations, and all these relations at higher redshifts are systematically lower than
the local \MZ\ relation of \citetalias{tre04} (dark blue solid curved line), once it has
been converted to the same metallicity calibration.\footnote{Again, we use the
  \cite{kew08} relation to transform to \citetalias{m91}-based metallicities.}
This result is consistent with previous studies of metallicity evolution with redshift.

\citet{sav05} used galaxies from the Gemini Deep Deep Survey and the Canada France
Redshift Survey to investigate the \MZ\ relation at $z=0.4$--0.98. A final sample of
56 galaxies was selected by the existence of rest-frame optical emission lines with
a $3\sigma$ detection limit of (0.6--3.2) $\times10^{-18}$ erg s$^{-1}$ cm$^{-2}$.
Metallicities were calculated using the $R_{23}$ line flux ratio and \citetalias{kk04}
calibration. The pink dashed-triple dot line in Figure \ref{fig:10} marks the
\citet{sav05} linear bisector fit, which does not agree with the NewH$\alpha$
  data (nor other higher redshift studies). However, the small sample size, lack of
selection criteria (i.e., no color selection, no S/N threshold for the $R_{23}$ emission
lines), and different fitting method for the \citet{sav05} data all prevent us from
directly comparing the two relations.

\citet[]{lam09} examined two subsets of $\sim$3000 $z=0.7$--0.9 galaxies from the VIMOS
VLT Deep Survey---a wide-shallow sample (6.1 deg$^{2}$ and $17.5\le I_{\rm AB} \le22.5$)
and a narrow-deep sample (0.61 deg$^{2}$ and $17.5\le I \le24$, green dot-dashed line
in Figure \ref{fig:10}).
Metallicities were calculated from rest-frame optical emission lines using the $R_{23}$
line flux ratio and the \citetalias{tre04} metallicity calibration.
Because the wide sample has shallower magnitude limits and is biased towards massive
galaxies, we choose to compare the NewH$\alpha$ \MZ\ relation against the deep sample
($N\sim40$ for $0.7<z<0.9$). Figure \ref{fig:10} shows that this sample is well within
the intrinsic scatter of the NewH$\alpha$ relation. \citet{lam09} observed that the
\MZ\ relation evolves towards lower overall metallicities more quickly for more massive
galaxies. However, low spectral resolution ($R_{s}\approx230$ or a large error domain
on AGN classification could also have an effect on both the \citet{lam09} samples; in
particular, contamination from AGN would falsely lower metallicity results, since
photoionization by AGN can produce rest-frame optical emission-line ratios that appear
similar to metal-poor star-forming galaxies \citep{mou11}.

\citet{cow08} used a sample of 154 galaxies from the 145 arcmin$^{2}$ GOODS-N field,
selected by rest-frame NIR bolometric flux (the limiting flux corresponding roughly to
an NIR magnitude of $K_s=23.4$). Metallicities were calculated using the $R_{23}$ line
ratio and both the \citetalias{kk04} and \citetalias{tre04} calibrations, although
equivalent widths (EWs) were used rather than emission-line fluxes. \citet{zah11}
noted a systematic error in the \citet{cow08} relation as a result of the fitting
procedure used---the $R_{23}$ metallicity diagnostic is less sensitive at the $R_{23}$
local maximum at $12+\log{\rm (O/H)}\sim8.4$, which leads to asymmetric metallicity
errors that influence the least-squares fit. We therefore follow the example of
\citet{zah11} and plot the median metallicities (red triangles in Figure \ref{fig:10})
rather than the functional form prescribed by the mean metallicities. The median
metallicities are within the scatter of the NewH$\alpha$ relation.

\citet{zah11} studied galaxies from the DEEP2 survey \citep{dav03}, which covers a
fairly large field of 3.5 deg$^{2}$. Out of 31,656 objects in the DEEP2 survey with
well-measured redshifts, only sources with spectra covering the wavelength range of
3720--5020 \AA\ (bracketing the rest-frame optical emission lines required for
$R_{23}$, see Section \ref{sec:metal}) and emission lines that could be fit were
considered in analysis. Further sample cuts included thresholds on H$\beta$ S/N,
H$\beta$ EW, combined $R_{23}$ errors, continuum fits, and the removal of sources
with $\log{R_{23}}>1$ (considered AGN), leaving a final sample size of
$\approx$1,600. As in \citet{cow08}, metallicities were calculated with the
$R_{23}$ ratio formula, using emission line EWs and the \citetalias{kk04} calibration.
Insufficient sample cuts for AGN, color contamination from red non-SF galaxies, and
a lack of S/N cut on \OII\ and \OIII\ lines would all tend to underestimate
metallicity, but we find that the \citet{zah11} relation (light blue dashed line in
Figure \ref{fig:10}) is extremely consistent with the NewH$\alpha$ data at overlapping
mass bins.

Finally, \citet{mou11} used a large ($\sim$3000 galaxies) sample from the AGN and Galaxy
Evolution Survey, which covers 7.9 deg$^2$. The survey is magnitude-limited
($I_{\rm AB}<20.45$), and emission-line galaxies were selected with the criteria that
H$\beta$ line flux is above $3\times10^{−17}$ erg s$^{-1}$ cm$^{-2}$ and log(\OII/H$\beta$)
$>$--0.3. Metallicities are computed using the EW formulas for the line ratios $R_{23}$ and
$O_{32}$ and the \citetalias{m91}, \citetalias{kk04}, and \citetalias{tre04} calibrations.
The \citet{mou11} and NewH$\alpha$ samples are complementary: the former is 
mass-incomplete below $10^{10.7}$ \Msun\ at $z\sim0.8$ as a result of the
survey flux limit, while the NewH$\alpha$ mass completeness drops off below
$\sim10^{9.5}$ \Msun\ because our data are emission-line selected. Therefore,
disregarding the NewH$\alpha$ data above an upper mass-completeness limit
($\sim$10$^{10.5}$ \Msun), the \citet{mou11} \MZ\ relation (orange diamonds in
Figure \ref{fig:10}) appears to be a smooth extension of the NewH$\alpha$
relation to higher masses.

\subsection{The Mass--Metallicity--SFR Relation}
\label{sec:mzs_relation}

Combining the stellar mass, metallicity, and SFR measurements for the New\Ha\ sample,
we now consider the correlation between all three derived properties.
As discussed in the introduction, residuals from the \MZ\ relation have been found to
correlate with the SFR in local galaxy samples. We aim to test whether a similar
secondary dependence on the SFR also exists in galaxies at $z\sim0.8$ as sampled by
the New\Ha\ dataset.

The New\Ha\ survey is one of the first surveys to offer both spectroscopic measurements
of the strong oxygen emission lines and \Ha\ narrowband fluxes at intermediate redshifts,
enabling more robust constraints on the nebular abundances and instantaneous SFRs.
These measurement methods are also commonly used at local redshifts, and allow a more
self-consistent test of whether the \MZS\ relation remains the same over cosmic time. 
To investigate the \MZS\ relation, we first remove all AGN, sources with poor SED fits
($\chi^{2}_{\nu}>10$), sources with an NB118 excess flux below 3$\sigma$, and limit
the dataset to the $R_{23}(3\sigma)$ sample. These restrictions remove 180 galaxies
  from the sample, leaving \Nthree\ galaxies for the following analysis. We note that the
  $R_{23}(3\sigma)$ cut removed the majority of galaxies (N=120).

Different parameterizations have been used to describe the local \MZS\ relation
\citepalias[see e.g.,][]{man10,lar10}. 
Some studies (\citetalias{lar13}; \citealt{hun12}) have assumed that this relation
can be accurately described by a plane.
In particular, \citetalias{lar13} argued that this may in fact be the best functional
representation of the \MZS\ relation, and refer to it as a ``fundamental plane.''
In contrast, \citetalias{man10} used a higher order parameterization (dubbed the
``fundamental metallicity relation''), which is motivated by the flattening at high
stellar masses in the \MZ\ relation (e.g., \citetalias{tre04}; \citealt{mou11}).
In addition, many studies \citep[e.g.,][]{ric11,xia12,bel13,sto13,hen13a,hen13b,zah13b,ly14,yab14}
have compared their samples against the fundamental metallicity relation to determine
if it holds at higher redshifts.
 
To enable direct comparison to \citetalias{lar10} and \citetalias{lar13}, we begin
by assuming that our relation can be accurately described by a plane.
Since commonly adopted methods for determining the best-fit plane may lead to different
results and interpretation, we use multiple techniques as described below: (1)
principal component analysis (PCA), (2) two-parameter regression, and (3) three-dimensional
$\chi^2$ minimization.

We then explore a higher order parameterization of the data following the methodology
of \citetalias{man10}. We later compare our analyses with previous works in Section
\ref{sec:disc}, and revisit the assumed plane parameterizaton in
Section~\ref{sec:MZSshape} and discuss possible implications.

We note that all the results that follow are based on metallicities determined
using the \citetalias{tre04} calibration, in order to facilitate direct comparison
with previous studies. Another metallicity calibration (\citetalias{m91} as used
previously for the \MZ\ relation) has been used, and we find that to first order, the
\MZS\ relation does not significantly differ.
The results of our planar fits with different metallicity calibrations
(\citetalias{tre04} and \citetalias{m91}) are summarized in Tables~\ref{table:PCA}, 
\ref{table:RegM}, and \ref{table:RegZ}.

\subsubsection{Principal Component Analysis}
\label{sec:PCA}

First, we conduct a PCA for our New\Ha\ dataset. This approach determines the eigenvectors
(called V1, V2, and V3), formed from linear combinations of the input parameters, that
are orthogonal to one another. 
One of the advantages of the technique is the ability to examine correlated measurements.
This is important for the \MZS\ relation, since the metallicity and SFR measurements are
strongly correlated with the derived stellar mass. Given these correlations, the
application of the PCA technique allows us to examine if a tilt in the direction of the
SFR is present for a plane parameterization.
This technique has been used by \citetalias{lar13} and \citet{hun12} to determine the
best-fit planar description for the \MZS\ relation.
We conduct our PCA analysis on the covariance matrix of our dataset with three variables:
\begin{eqnarray}
  \nonumber
  x_1 & \equiv & \log{\left(\frac{M_{\star}}{M_{\sun}}\right)},\\
  x_2 & \equiv & 12+\log({\rm O/H}),{\rm~and}\\
  \nonumber
  x_3 & \equiv & \log{\left[\frac{\SFRHa}{M_{\sun}~{\rm yr}^{-1}}\right]}.
\end{eqnarray}
To account for measurement uncertainties in the PCA, we conduct Monte Carlo realizations
of our data, where the stellar mass, SFR, and oxygen abundance for each galaxy in the
sample are drawn 100,000 times from a Gaussian probability distribution defined by the
1$\sigma$ errors in each parameter.
We then fit each simulated sample of \Nthree\ galaxies using the PCA code available
through the NASA IDL Astronomy User's Library.

\begin{figure}
  \epsscale{1.1}
  \plotone{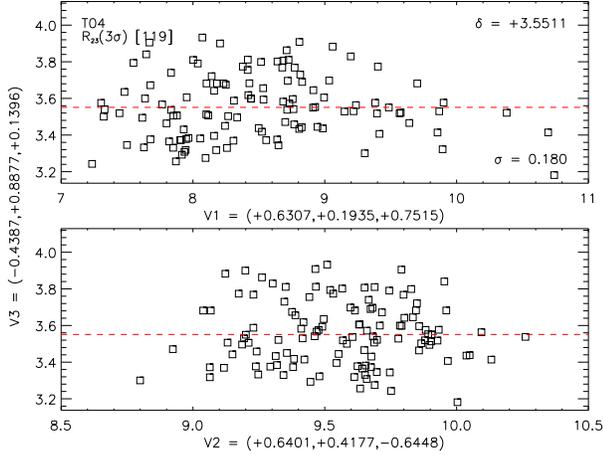}
  \caption{PCA of our $R_{23}(3\sigma)$ sample with metallicities determined using the
    \citetalias{tre04} calibration. The first and second principal components (V1 and V2)
    are shown in the top and bottom panels, compared with the third component (V3) with
    the least amount of variance. The dashed line represents the average of V3. Our sample
    can be described by Equation~(\ref{eqn:PCA_FP}) with ($\alpha$,$\beta$,$\gamma$)
    $\approx$ (--0.384, 0.923, 0.087). The PCA results from our Monte Carlo realizations
    are provided in Figure~\ref{fig:12}.}
  \label{fig:11}
\end{figure}

We find that the first two principal components account for 78.8\%$\pm$1.6\% and
15.2\%$\pm$1.5\% of the variance, respectively.
The first principal component, which has the largest variance, is
V1 = (0.610, 0.183, 0.771). The other two eigenvectors are V2 = (0.690, 0.352, --0.629),
and V3 = (--0.384, 0.923, 0.087).
In Figure~\ref{fig:11}, the data are projected in the planes defined by these
principal components.

\begin{figure*}
  \epsscale{1.1}
  \plotone{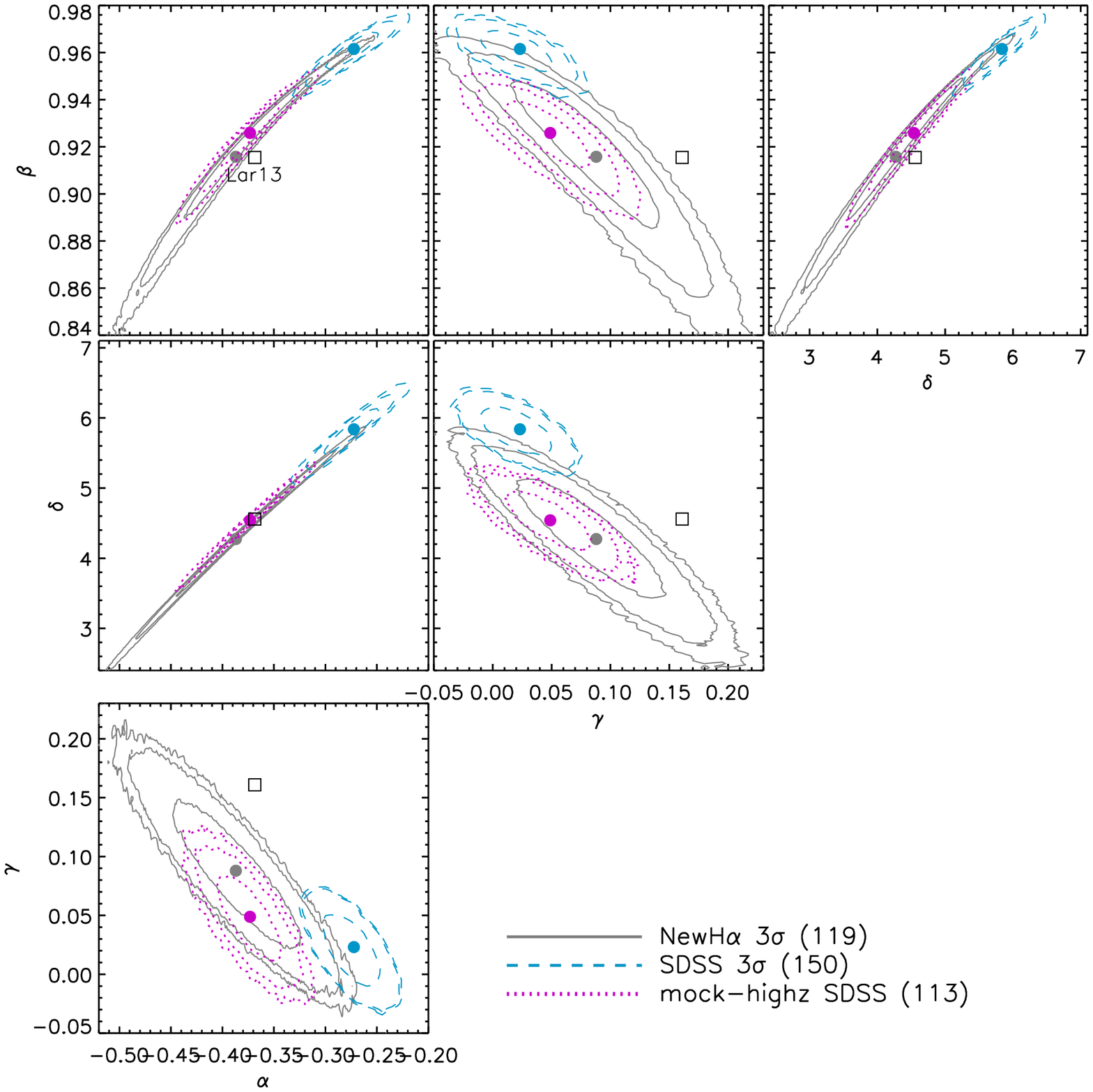}
  \caption{Results of Monte Carlo PCA fitting to the New\Ha\ $R_{23}(3\sigma)$ sample of
    \Nthree\ galaxies (grey solid lines) compared to 150 randomly-selected local galaxies
    from the SDSS (blue dashed lines). We also overlay a sample (``mock-highz SDSS'';
    see Section~\ref{sec:MZSsample}) selected from the SDSS to have similar stellar
    masses and SFRs to New\Ha\ galaxies (purple dotted lines). Coefficients $\alpha$,
    $\beta$, $\gamma$, and $\delta$ describe a plane as given in Equation
    (\ref{eqn:PCA_FP}). Contours at 68\%, 95\%, and 99\% confidence are shown.  These
    contours demonstrate that the best-fitting PCA plane for the New\Ha\ sample is
    consistent with fits to SDSS galaxy samples with similar sample size. The standard
    PCA fit of \citetalias{lar13} is shown by the open squares.}
  \label{fig:12}
\end{figure*}

Since V1 and V2 have the largest dispersions, they can be interpreted as vectors that
lie along the best-fit plane, while V3 is the vector that is orthogonal to the plane.
Figure \ref{fig:11} illustrates that V3 has the least amount of variance with an rms
of $\approx$0.18 dex. This low dispersion is critical, as it suggests that V3 provides
a mathematical description for the best-fit plane such that a combination of stellar
mass, SFR, and metallicity yields a constant:
\begin{equation}
  \alpha x_1 + \beta x_2 + \gamma x_3 = \delta,
  \label{eqn:PCA_FP}
\end{equation}
with ($\alpha$,$\beta$,$\gamma$,$\delta$) = (--0.384$^{+0.03}_{-0.04}$, 0.923$\pm$0.02,
0.087$^{+0.04}_{-0.03}$, +4.301$^{+0.57}_{-0.46}$).
The V3 results from our Monte Carlo realization are shown in Figure \ref{fig:12}.
A summary of our PCA results, using different metallicity calibrations
\citepalias{m91,tre04} and sample selections, can be found in Table
\ref{table:PCA}.

Assuming that our data can be described by a plane, $\gamma$ can be interpreted to
signify the importance of the SFR in the correlation. The PCA shows that $\gamma$ is
non-zero ($\approx$3$\sigma$ significance), suggesting that the plane which best
describes our dataset is moderately tilted in the SFR dimension.

\subsubsection{Two-Parameter Regression}
\label{sec:reg}
\begin{figure*}
  \epsscale{1.1}
  \plotone{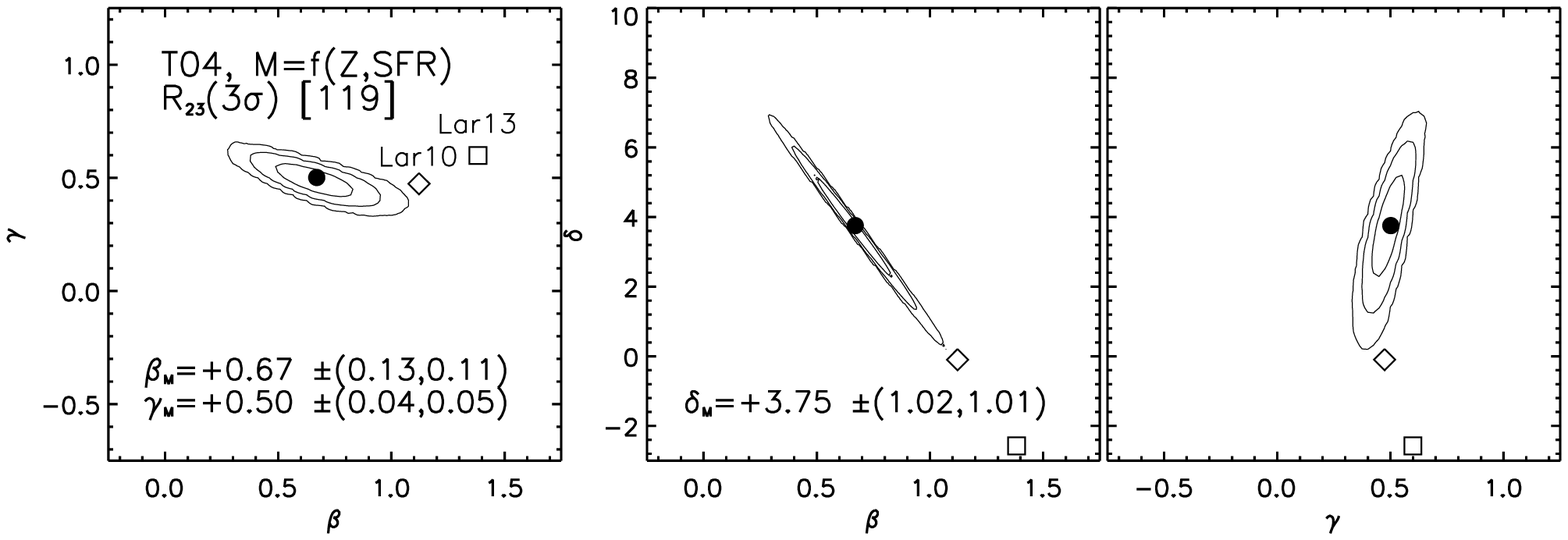}
  \plotone{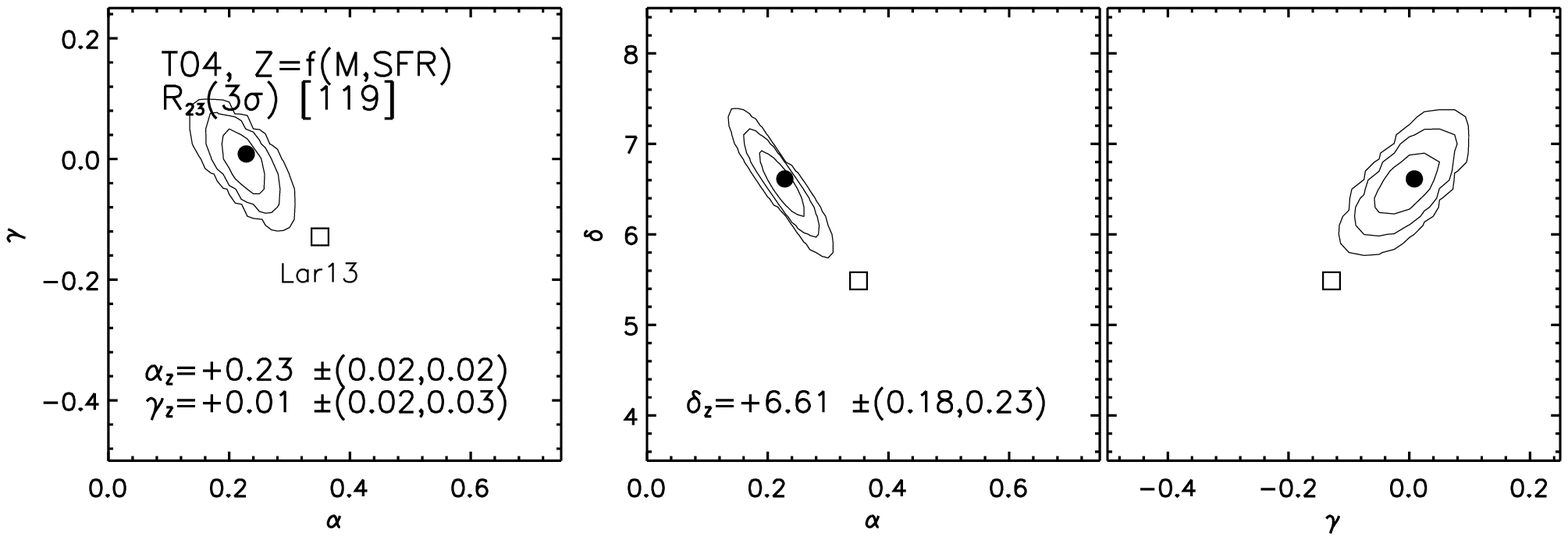}
  \caption{Results of our Monte Carlo simulations for regression-based plane fitting.
    The top panels show the fitting conducted with stellar mass as the dependent
    variable, $M=f(Z,{\rm SFR})$, while the bottom panels consider $Z=f(M,{\rm SFR})$.
    Metallicities are based on the \citetalias{tre04} calibration. The regression fits
    of \citetalias{lar10} and \citetalias{lar13} are shown as open diamonds and
    squares, respectively.}
  \label{fig:13}
\end{figure*}

Another approach for finding the best-fit plane is linear regression, whereby one 
parameter is modeled in terms of two other parameters. We consider a plane
parameterization used to describe local galaxies \citepalias{lar10}, where the
stellar mass is treated as the dependent variable:
\begin{equation}
  \log{\left(\frac{M_{\star}}{M_{\sun}}\right)} = \beta_{M}Z + 
  \gamma_M\log{\left(\frac{\SFRHa}{M_{\sun}~{\rm yr}^{-1}}\right)}+\delta_M,
\label{eqn:M_FP}
\end{equation}
where $Z \equiv 12+\log{\rm (O/H)}$. We conduct the regression using the IDL routine
MPFIT \citep{mar09}, which uses the Levenberg-Marquardt least-squares minimization
technique.
As with the PCA, we also perform a Monte Carlo simulation to determine the uncertainties
in the fit. Here, the Gaussian randomization only occurs in the two independent variables
(SFR and metallicity) and measured uncertainties in the dependent variable (stellar mass)
are accounted for in the least-squares minimization.
The results of our Monte Carlo simulation are shown in Figure \ref{fig:13}, and are
reported in Table~\ref{table:RegM} using two different metallicity calibrations
  \citepalias{tre04,m91}.
The best fitting parameters that describe the New\Ha\ sample are
$\beta_M=0.67^{+0.13}_{-0.11}$, $\gamma_M=0.50^{+0.04}_{-0.05}$, and
$\delta_M=3.75^{+1.02}_{-1.01}$.

While Equation~(\ref{eqn:M_FP}) was reported to yield the lowest $\chi^2$ in all three
observables for local SDSS galaxies \citepalias{lar13}, it can be viewed as
counter-intuitive\footnote{Other galaxy properties have been extensively compared
  against the stellar mass (i.e., the latter is treated as the independent variable).}.
A better description of the plane, which is a simple extension of the \MZ\ relation, is:
\begin{equation}
  Z = \alpha_Z\log{\left(\frac{M_{\star}}{M_{\sun}}\right)} +
  \gamma_Z\log{\left(\frac{\SFRHa}{M_{\sun}~{\rm yr}^{-1}}\right)}+\delta_Z.
\label{eqn:fp2}
\end{equation}
The regression fitting for our sample using this projection, as shown in
Figure \ref{fig:13}, yielded $\alpha_Z=0.23\pm0.02$, $\gamma_Z=0.01^{+0.02}_{-0.03}$, and
$\delta_Z=6.61^{+0.18}_{-0.12}$ for \citetalias{tre04} metallicities.
The results of our $Z=f(M,{\rm SFR})$ regression are consistent with our previous \MZ\
least-squares fitting, which found a slope ($\alpha_{Z}$) of 0.25 and a constant
offset ($\delta_{Z}$) of 6.23 (Section~\ref{sec:MZR}). This regression analysis
demonstrates that a strong secondary dependence on the SFR for a \MZS\ plane is
\emph{not} present in our dataset.
We summarize our $Z=f(M,{\rm SFR})$ regression results for \citetalias{tre04}
and \citetalias{m91} in Table \ref{table:RegZ}.

\subsubsection{Three-Dimensional $\chi^2$ Minimization}
\label{sec:chi2}
One of the limitations of the two-parameter regression approach---used to study the SDSS
sample by \citetalias{lar10} and \citetalias{lar13}---is the arbitrary choice of the
independent and dependent variables, as we have discussed.
To address this, we consider a three-parameter fit that simultaneously minimizes $\chi^2$
in all three dimensions. We assume that the data can be described by a plane as given
by Equation~(\ref{eqn:PCA_FP}).
This method complements the PCA, since it is less susceptible to outliers that directly
affect the covariance matrix, and hence the principal components (see Section 
\ref{sec:MZSplanefit} for further discussion). It is a three-dimensional extension of
the $\chi^2$ estimator used by \citet{tre02}, for example. 
To obtain meaningful errors, we scale our measurement uncertainties to yield a reduced
$\chi^2$ of 1.
We find a best fit of ($\alpha$,$\beta$,$\gamma$,$\delta$) =
(--0.37$\pm$0.05, 0.92$\pm$0.02, 0.10$\pm$0.05, 4.54$\pm$0.64).
These values are similar to those determined from PCA, again suggesting that there is at
most a moderate dependence of metallicity on the SFR for \Ha-selected galaxies from the
New\Ha\ survey.

\subsubsection{Non-planar Formalism of \cite{man10}}
\label{sec:nonlinear}
As previously stated, a curved-surface parameterization may be a better representation
of the \MZS\ relation. With this in mind, we split our sample into low-mass and
high-mass subsamples and perform PCA. The results are summarized in Table~\ref{table:PCA}.
We find that the planes which best fit the low-mass sample are significantly different
from those that best fit the high-mass sample.
  
We therefore consider a non-planar fit between the three derived properties
following \citetalias{man10}, who determined a curved-surface representation of the
\MZS\ relation at local redshifts.
\citetalias{man10} calculated metallicities for local SDSS galaxies with
two separate emission-line flux ratio measurements---the
\cite{n06} \NII\,$\lambda$6583/\Ha\ calibration and the \citet{mai08} $R_{23}$
calibration. In cases where both measurements agree within 0.25 dex, an average
of the two was used. Since \NII/\Ha\ measurements do not exist for our sample, direct
comparison of the New\Ha\ sample to \citetalias{man10} is difficult. In particular, while
we could examine whether the \MZS\ relation exists in our sample using {\it only} 
$R_{23}$-based metallicities estimated from the \citet{mai08} calibration, we note that
this approach has yet to be conducted with SDSS galaxies (i.e.,
excluding \NII/\Ha\ measurements). A \citet{mai08} $R_{23}$-based
\MZS\ relation is beyond the scope of this paper. Therefore, we choose to
compare to the relation determined by \citetalias{yat12}, who followed the same procedure as
\citetalias{man10} but used \citetalias{tre04} metallicities.
  
We determine a second-order polynomial fit of metallicity as a function of a
linear combination of mass and SFR:
\begin{equation}
  \mu_{\alpha} = \log{\left(\frac{M_{\star}}{M_{\sun}}\right)}-
  \alpha \log{\left(\frac{{\rm SFR}}{M_{\sun}~{\rm yr}^{-1}}\right)},
\end{equation}
where $\alpha$ is a free parameter chosen to minimize the scatter of metallicity. Here,
$\alpha=0$ corresponds to the \MZ\ relation while $\alpha=1$ refers to metallicity
having an inverse dependence with the specific SFR.

\begin{figure}
 \epsscale{1.1}
  \plotone{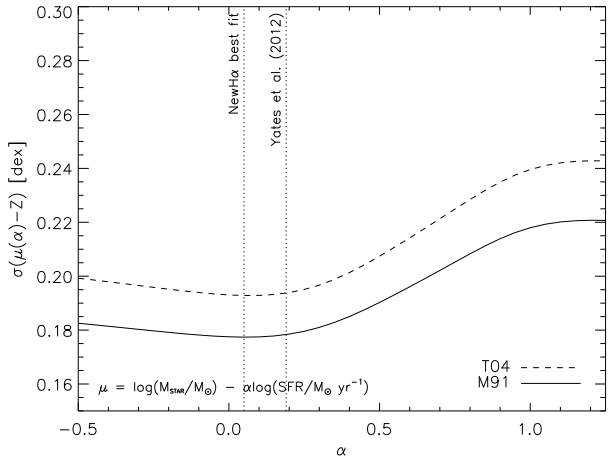}
 \caption{Dispersion in oxygen abundances for different projections of the \MZS\ relation
   using a curved-surface least-squares fitting approach with
   $\mu=\log{\left(\frac{M_{\star}}{M_{\sun}}\right)}-
   \alpha \log{\left(\frac{{\rm SFR}}{M_{\sun}~{\rm yr}^{-1}}\right)}$.
   Calculations were conducted using \citetalias{m91}- (solid line) and
   \citetalias{tre04}-based (dashed line) metallicities. The analyses show that a strong
   secondary dependence on SFR does not exist, consistent with planar fitting approaches;
   however, the data cannot exclude or distinguish between zero and weak dependence
   \citepalias[$\alpha=0.19$;][]{yat12}.}
   \label{fig:14}
\end{figure}

\begin{figure*}
 \epsscale{1.1}
  \plottwo{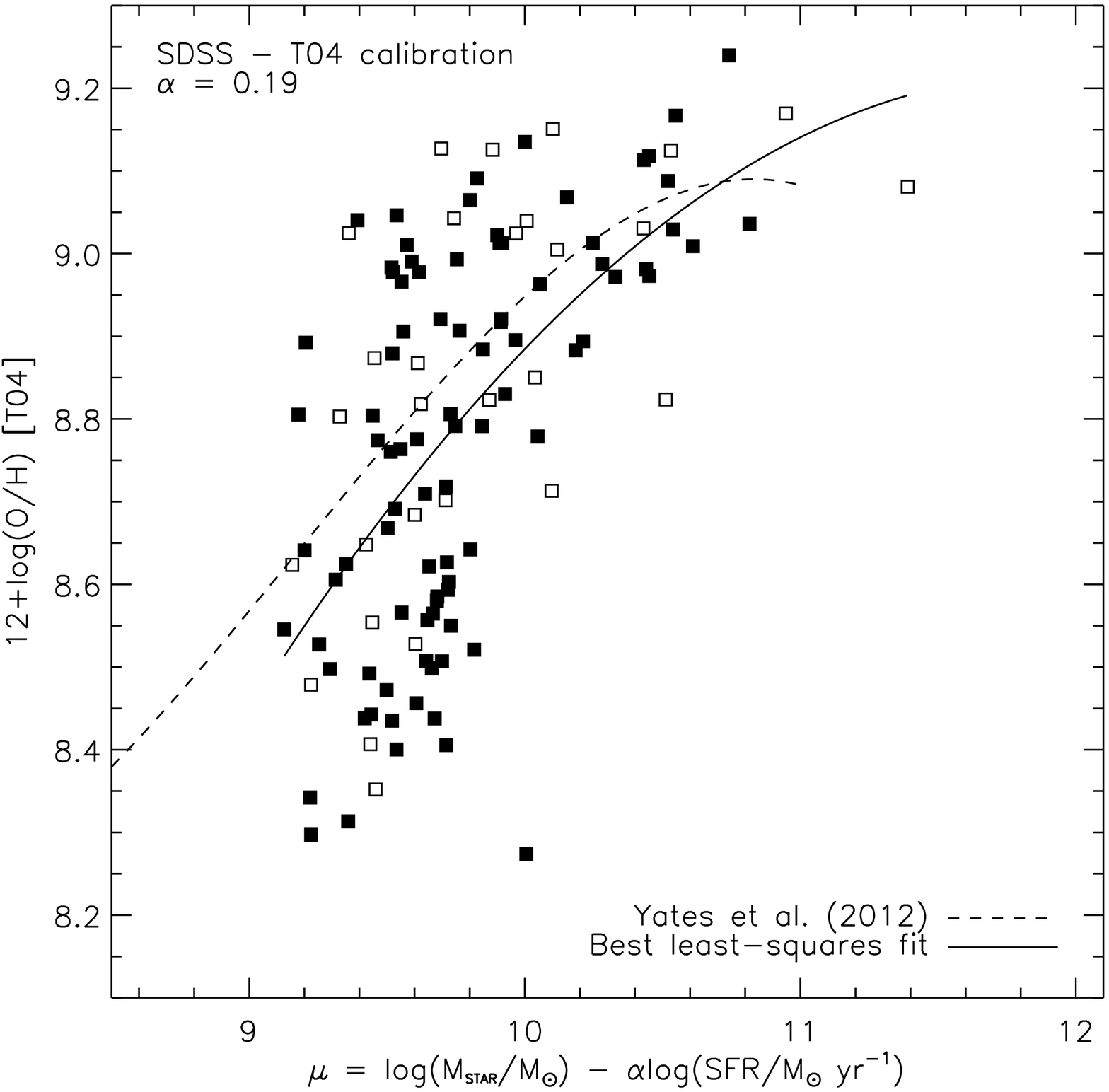}{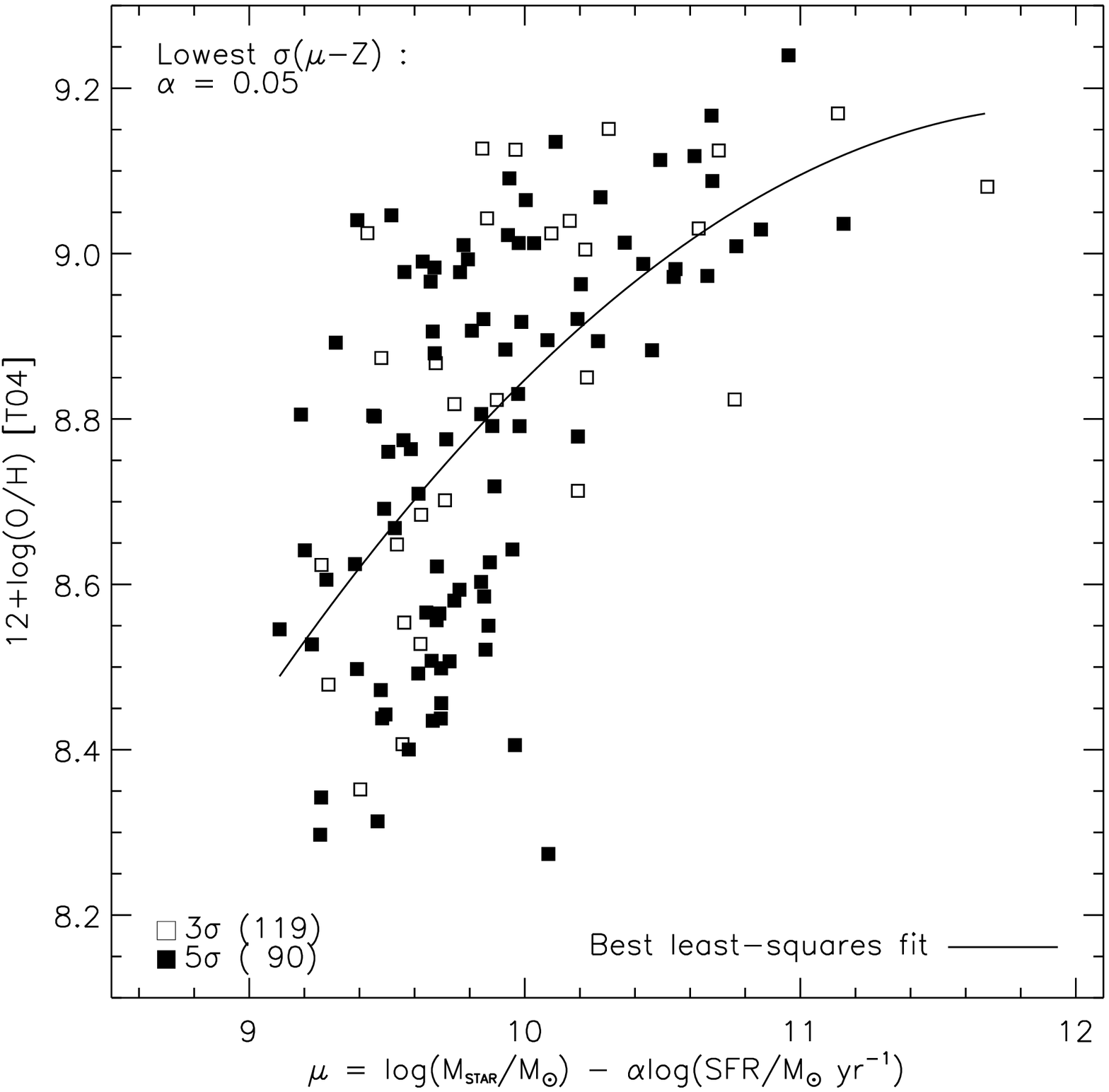}
  \caption{\MZS\ relation following the projection ($\mu$--$Z$) of \citetalias{man10}.
    Here we illustrate best-fit projection for local SDSS galaxies
    \citepalias[left, $\alpha=0.19$;][]{yat12} and the projection with the lowest
    dispersion for the New\Ha\ sample (right, $\alpha=0.05$). The best-fitting third-order
    polynomial of \citetalias{yat12} is shown by the dashed line while our best fits
    are shown by the solid lines. Filled squares show the $R_{23}(5\sigma)$ sample with
    additional sources in the $R_{23}(3\sigma)$ sample (opened squares). For direct
    comparisons to \citetalias{yat12}, we use \citetalias{tre04}-based metallicities.}
  \label{fig:15}
\end{figure*}

We consider a range of $\alpha$ values, and illustrate in Figure \ref{fig:14} 
the dispersion of the best-fit second-order polynomial. This result demonstrates that
scatter in metallicity is minimized at $\alpha\sim0.05$ (i.e., suggesting weak
dependence on the SFR). High values of $\alpha$ ($\gtrsim0.5$) can be excluded,
suggesting that a strong dependence on SFR does not exist. However, we cannot exclude
moderate dependence \citepalias[e.g., $\alpha=0.19$ for local galaxies as
determined by][]{yat12}, since the scatter in metallicity does not significantly
change for $\alpha\lesssim0.5$.
This result is illustrated in Figure \ref{fig:15}, where we plot the best-fit
second-order polynomial for both $\alpha=0.05$ and $\alpha=0.19$.

\section{Discussion}
\label{sec:disc}
As discussed in the introduction, the detailed relation between galaxy stellar mass,
SFR, and gas-phase metallicity is important for understanding inflows and outflows of
gas, and the chemical evolution of galaxies. The shape of this relation and the degree
to which it does or does not evolve with redshift can provide insights into whether the
processes governing the interaction between galaxies and their surrounding medium are
``fundamental'' in the sense that they may not differ substantially at various points
in cosmic time.

Several studies \citepalias[e.g.][]{man10,lar10,lar13} have reported that along with
the well-established strong correlation between mass and metallicity, there is a
moderate, but significant correlation with the SFR. Furthermore, a number of studies,
including \citetalias{man10} and \citet{hun12}, have indeed found that galaxies at
redshifts up to $z\sim3$ can be described by the same \MZS\ relation. However,
contradictory results have recently been reported: \citet{san13} and \citet{hug13} were
unable to find a significant correlation with the SFR, and argue that previous results
based on the SDSS dataset were spurious due to aperture effects
\citep[for further discussion, see Section 4 of][]{san13}. Also, \citet{zah13b} found
evidence for redshift evolution in the \MZS\ relation.

In Section \ref{sec:mzs_relation} of this paper, we examined if the \MZS\ relation at
$z\sim0.8$ exists using the NewH$\alpha$ dataset---and if it does, to determine if it
is consistent with previous analyses of local galaxies from the SDSS dataset. 
To facilitate comparison to local results, we followed previous approaches
by assuming that the \MZS\ relation can be described by a plane or a surface.
For the planar description, we used PCA, two-parameter regression, and $\chi^2$
minimization with metallicities calibrated against \citetalias{tre04}.
We found that the NewH$\alpha$ data show a moderate dependence ($\gamma\approx0.1$) of
the \MZ\ relation on SFR. This is slightly lower than the dependence found in the local 
universe ($\gamma\approx0.16$; \citetalias{lar13}).
For the curved-surface parameterization, we use least squares fitting to describe a
second-order polynomial between metallicity and a combination of mass and SFR. This
analysis excludes a strong dependence of the \MZ\ relation on SFR; however, it cannot
distinguish between moderate and no dependence.

How do we interpret our results, and where do they fit in within the current debate on
the \MZS\ relation?  We address these questions by first investigating whether some
limitation(s) of our analyses or dataset may obscure the true underlying relationship.
We ask:
\begin{enumerate}
\item Is our result biased or affected by some limitation of the NewH$\alpha$ dataset? 
  (Section~\ref{sec:MZSsample})
\item Is our result biased or affected by the chosen plane-fitting techniques?
  (Section~\ref{sec:MZSplanefit})
\item Finally, is our result biased or affected by our assumed parameterizations of
  the dataset? (Section~\ref{sec:MZSshape})
\end{enumerate}

\subsection{Limitations of the NewH$\alpha$ Dataset}
\label{sec:MZSsample}

The following sample limitations may, individually or in combination, bias our measurement
of the SFR dependence of the \MZS\ relation: (1) small sample size, (2) measurement
uncertainties, and (3) restricted coverage of parameter space. These limitations apply
generally to any study attempting to construct a \MZS\ relation. 

In this work, we focus on the first possible limitation: the small size of the NewH$\alpha$
sample (119 galaxies) used for studying the \MZS\ relation. To understand the effects of
sample size we construct ``mock'' samples from the SDSS DR7 sample. Here, the MPA-JHU
catalog provides total stellar masses from fitting the
$u\arcmin g\arcmin r\arcmin i\arcmin z\arcmin$ photometry \citep{sal07}, total SFRs
primarily from Balmer emission lines \citep{bri04}, and metallicity within the optical
fibers following \citetalias{tre04}. Restricting our sample to
galaxies\footnote{Selected by the \cite{bal81} diagnostic selection following \cite{kau03}
  with at least a 3$\sigma$ detection for \Ha, \NII, \OIII\ $\lambda$5007, and \Hb.}
with estimates of stellar mass, metallicity, and SFR, and redshift between $z=0.07$ and
$z=0.30$, we have a working SDSS sample of 90,686 galaxies.

We therefore begin with a randomly-selected subsample of 150 galaxies and then consider
improvements to our base sample by increasing the sample size. We fit each SDSS subsample
with a plane using the PCA technique, as discussed in Section \ref{sec:PCA}, for direct
comparisons with our New\Ha\ results. For the smallest subsample, we find little
dependence on the SFR, with $\gamma\approx0.02$.
We then increase the sample size in increments of 150 galaxies, with the expectation that
the larger sample size will provide more definitive constraints on SFR dependence. However,
even with a sample of 1950 galaxies (13 times larger than the base sample), $\gamma$
remains at or below 0.05. Extending the PCA analysis to the largest sample possible
(N = 90,686), we find that a surprisingly weak dependence exists on the SFR ($\gamma=0.02$)
for a planar description of the \MZS\ relation.
This dependence is roughly three to eight times weaker than that found by
\citetalias{lar13}, but consistent (within errors) with results from the analysis based
on the NewH$\alpha$ dataset reported here ($\gamma\approx0.087^{+0.04}_{-0.03}$). This is
best demonstrated in Figure~\ref{fig:12} where the mock SDSS sample, with similar sample
size (N = 150) to the New\Ha\ sample, is shown by the dashed (blue) line contours.

How can the results of our experiments with mock SDSS samples be reconciled with the
results of \citetalias{lar13}? The sample used by \citetalias{lar13} used stricter
selection cuts. Their strictest constraint is on the S/N for the \OIII\ $\lambda$5007,
\Hb, \Ha, and \NII\ emission lines (requiring at least 8$\sigma$).
The \Ha\ restriction biases their sample toward higher SFRs, while the \OIII\ restriction
preferentially selects against metal-rich galaxies, leaving metal-poor galaxies with
higher SFRs in the sample.
Interestingly, we do find that a S/N restriction of 8 on the nebular emission lines
yielded a higher SFR coefficient ($\gamma\approx0.08$--0.14 with the standard PCA) based
on sample sizes that span 150 to 31,477 galaxies.

The mock SDSS samples used in the analysis described in this section, on the other hand,
were constrained by similar emission-line restrictions ($3\sigma$ detection) used for
the NewH$\alpha$ dataset, and achieved results more consistent with the NewH$\alpha$
results.
To further demonstrate this point, we also selected from the SDSS galaxies with similar
($\leq$0.1 dex) stellar masses and SFRs to those in the New\Ha\ sample (hereafter
``mock-highz SDSS''). Because local galaxies have lower SFRs, six of \Nthree\ New\Ha\
galaxies do not have a ``local analog.'' The results of the PCA for the mock-highz SDSS
sample are shown as dotted (purple) line contours in Figure~\ref{fig:12}, and are in
better agreement with the New\Ha\ PCA results.
These comparison results suggest that the differences in sample selection may therefore
produce the observed discrepancy between the results of NewH$\alpha$ and the results
of \citetalias{lar13}.

\subsection{Limitations of Principal Component Analysis}
\label{sec:MZSplanefit}

Another potential issue with our investigation is our reliance on the PCA technique to
find the plane that best describes the \MZS\ relation, and to compare to results based
on local galaxy samples. This technique has shortcomings, particularly in its sensitivity
to outliers.

PCA finds eigenvectors (or principal components) formed from linear combinations of input
parameters ($M_{*}$, $Z$, and SFR). Since variance-dependent calculations are used to
determine the principal components, outliers may strongly skew PCA results. Considering
the uncertainties of derived quantities and large size of the SDSS sample, there are
significant numbers of (true) outliers in the sample that would suggest that the PCA
technique is unreliable for the SDSS.

\begin{figure}
 \epsscale{1.1}
  \plotone{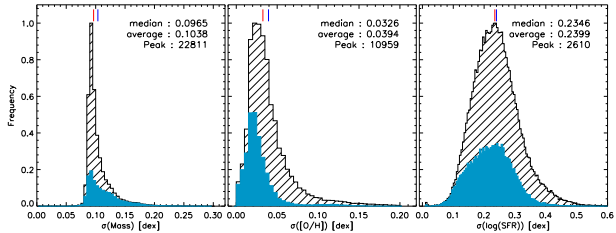}
 \caption{Distribution of measurement uncertainties for stellar mass (left), metallicity
   (middle) and SFR (right) of the SDSS sample. The full sample of 90,686 galaxies is
   shown cross-hatched, while a smaller sample with restrictions on the \OIII, \Hb, \Ha,
   and \NII\ lines (S/N $\ge$ 8) is shown filled-in (blue in online version).}
  \label{fig:16}
\end{figure}

This is particularly demonstrated when we account for measurement uncertainties through 
Monte Carlo techniques in the PCA fitting (see Section \ref{sec:PCA}). We find a different
best-fit plane (albeit one that still has a low $\gamma$) when compared to a standard
PCA (i.e., without considering uncertainties). This is expected because of the
uncertainties on the SFRs: at least 32\% of the SDSS sample deviates significantly
($\geq$0.22 dex; see Figure \ref{fig:16}) from what is likely the best-fitting plane.
When performing the same analysis with an SDSS sample similar to that of
\citetalias{lar13}, we also find a different result for $\gamma$, $\sim$0.12 versus the
reported result of 0.16.

We suggest that instead of PCA, three-dimensional $\chi^2$ minimization (see Section
\ref{sec:chi2}) should be the preferred method of parameterizing the \MZS\ relation as
a plane.  
The three-dimensional $\chi^2$ minimization technique fits all three observables
simultaneously and is less susceptible to outliers: galaxies with more uncertain 
measurements are downweighted relative to those with more precise measurements. In the
case of the NewH$\alpha$ dataset, our results are consistent between PCA and
three-dimensional $\chi^2$ fitting, suggesting that our sample is not as severely
affected by the PCA analysis.
Nevertheless, we recommend caution when proceeding with PCA analysis without understanding
the effects of outliers. 

\subsection{Limitations of the Parameterization of the \MZS\ Relation}
\label{sec:MZSshape}
In much of our analysis---including our investigation of potential sample size
and PCA technique limitations---we have adopted a plane to describe our data.
However, as noted in Section \ref{sec:nonlinear}, this assumption may be wrong.
If there is in fact a \MZS\ relation which is fundamental (i.e., universally describes
galaxies at all redshifts), and there is curvature in that relation, studies which
assume a plane parameterization in their analysis may mistakenly infer evolution in the
relation. That is, evolution in a planar relation may actually be a result of sampling
different parts of the curved surface relation with respect to redshift. At $z\sim0.8$,
the NewH$\alpha$ sample has lower average metallicity and higher average SFR than the
local SDSS sample does. 
Furthermore, if the \MZS\ relation is curved, the results from a plane fit can be
different if the sample is limited in parameter space.
This is demonstrated in Section~\ref{sec:nonlinear} when we split our sample
into low-mass and high-mass subsamples. The plane that best fits the low-mass sample
is significantly different from the one that fits the high-mass sample. This implies
that our sample follows a non-planar projection---a discrepancy is unsurprising if
there is no truly ``good'' planar fit.

With this in mind, we follow the procedure of \citetalias{man10} and
\citetalias{yat12} to find the projection of least scatter, as described in Section
\ref{sec:nonlinear}. In this projection, the parameter $\alpha$ describes the
dependence of the \MZ\ relation on the SFR. Our dataset excludes a strong dependence
on SFR ($\alpha\gtrsim0.5$); however, it cannot distinguish between moderate
($\alpha\sim0.2$) and no dependence. This result is consistent with those reported
for local galaxies \citepalias[$\alpha=0.19$;][]{yat12}.

We note that although several \MZS\ studies have followed the methods of \citetalias{man10}
\citep{yat12, and13}, the effects of binning the SDSS sample by both mass and SFR have
been a point of some contention, as \citetalias{lar13} have argued that the grid adopted
for data binning can effectively change the shape of the curved-surface \MZS\ relation. 
In addition, this method of projection of least scatter relies on an initial
assumption of a polynomial functional form.

We therefore suggest that future work be done to investigate \emph{non-parametric}
methods of fitting the \MZS\ relation. For instance, the Kolmogorov-Smirnov (K-S) test
is used to compare a one-dimensional sample with a reference probability distribution. 
An extension of the K-S test to three dimensions would be ideal for fitting the \MZS\
relation while avoiding assumptions about the shape or functional form of the relation.

\section{Conclusions}
\label{sec:concl}

We have studied the relationships between stellar mass, SFR, and metallicity using a
sample of \Nspec\ galaxies at $z\sim0.8$ selected by the presence of \Ha\ emission in a
narrow bandpass filter. Deep optical spectra obtained with Magellan IMACS enable us to
measure gas-phase metal abundances with various theoretical and empirical oxygen-based
calibrations, and to compare them to SFRs estimated from the \Ha\ luminosity and stellar
masses from SED modeling.

Our emission-line galaxy sample spans stellar masses from $\sim$10$^9$ to $6\times10^{11}$
\Msun, H$\alpha$-based SFRs between 0.4 and 270 \Msun\ yr$^{-1}$, and metallicities from
$12+\log{({\rm O/H})}=8.3$ to 9.1 ($Z/Z_{\sun} = 0.4$--2.6) on a metallicity scale based
on the \citetalias{m91} calibration.

We compared H$\alpha$-based SFRs with SFRs estimated from SED fitting (i.e. FUV-based SFRs).
We found that once both measures were corrected for dust attenuation with optical depths
computed from SEDs, the two measures agreed well (median offset of $\sim$0.09 dex) with
low dispersion ($\sim0.2$ dex). In addition, this agreement  holds for the full range of
stellar mass and for high SFRs ($\gtrsim$100 \Msun\ yr$^{-1}$).

Based on a linear least-squares fit over stellar masses between $10^{9.1}$ \Msun and
$10^{11.7}$ \Msun, the \MZ\ relation for our sample is $12+\log{({\rm O/H})} =
(0.25\pm0.03)\log{\left(\frac{M_{\star}}{M_{\odot}}\right)}+(6.23\pm0.33)$.
This is consistent with previously reported results for galaxy samples at similar
redshifts. At fixed stellar mass, the \MZ\ relation for our sample is systematically
lower by $0.1$ dex in metallicity than the local SDSS relation of \citetalias{tre04}. 

Similarly, we found a NewH$\alpha$ \MSFR\ relation of
$\log{\left(\frac{\SFRHa}{M_{\odot}~{\rm yr}^{-1}}\right)} =
(0.75\pm0.07)\log{\left(\frac{M_{\star}}{M_{\odot}}\right)}-(6.73\pm0.67)$, 
which is consistent with literature results at similar redshifts (within 0.15 dex in
SFR of previous results). This consistency is somewhat surprising given that the 
NewH$\alpha$ sample is \Ha\ selected, which might bias our relation toward higher SFR. 
However, this suggests that our sample is in fact relatively complete down to low EWs. 

We then calculated the best-fit plane describing the stellar masses, SFRs, and
metallicities of the NewH$\alpha$ sample using three methods: principal component 
analysis, two-parameter regression, and three-dimensional $\chi^2$ minimization. 
The fits resulting from all these analyses at $z\sim0.8$ showed only a moderate secondary
dependence on the SFR weaker than that reported by \citetalias{lar10} and \citetalias{lar13}.
In addition, we considered a curved-surface parameterization following \citetalias{man10},
and found that the New\Ha\ sample is consistent with local studies \citepalias[i.e., a 
weak dependence of the \MZ\ relation on SFR;][]{yat12}, and excludes a strong
SFR dependence.

To better understand the possible implications of these results, we asked whether some
limitation of our dataset and/or analysis may obscure a stronger or weaker dependence 
on the SFR by using mock samples drawn from the SDSS. 

We started by examining possible issues associated with the small size of our sample. Using
a randomly-selected subsample of 150 SDSS DR7 galaxies using the PCA technique, we found a
dependence on the SFR that was three to eight times weaker than the SDSS study of
\citetalias{lar13}.
Somewhat surprisingly however, increasing the sample size did not significantly change 
this result, even using the largest possible sample (N$\approx$90,000). We learned that
differences in the adopted signal-to-noise cuts may lead to apparently significant
differences in the level of the second parameter dependence on the SFR. By imposing cuts
on the mock sample that were more similar to the ones used to form the NewH$\alpha$
dataset, we found a weaker SFR dependence more consistent with the one reported here.
Further work is needed to reconcile these results with recent studies based on IFU and
drift-scan observations of local galaxies which find that there is no secondary dependence
on the SFR \citep{san13,hug13}.

We also examined potential issues in our fitting analysis, and the lessons learned here
are of use for future \MZS\ studies. For example, we find that the PCA technique is
highly sensitive to outliers and measurement uncertainties, and three-dimensional
$\chi^{2}$ minimization may be preferred as a more robust plane-fitting technique. This
is particularly true for the sample size analysis described above, as there are
significant numbers of true outliers in the SDSS dataset.

We conclude that future work should include the following. Locally, the SDSS galaxies
excluded by the \citetalias{lar13} analysis should be examined more closely, and
potential systematics in SFR and $Z$ measurements due to SDSS aperture effects can be
verified directly with forthcoming integral field spectroscopic surveys (e.g.,
MaNGA and SAMI). Future works, particularly those based on higher redshift samples, 
should also account for dataset limitations in constraining a possible weak secondary
dependence in the SFR. Here, we have addressed the effects of small sample size, but
limited coverage of parameter space and relatively large measurement uncertainties may
also have biasing effects.
Finally, we stress the need for a non-parametric method of fitting a three-dimensional
dataset in order to truly determine an \MZS\ relation without making assumptions about
the shape or functional form of the relation.

\acknowledgements
This work is based on observations obtained with MegaPrime/MegaCam, a joint project of 
CFHT and CEA/DAPNIA, at the CFHT, which is operated by the National Research Council
of Canada, the Institut National des Science de l'Univers of the Centre National de la
Recherche Scientifique of France, and the University of Hawaii.
We thank Sebastien Foucaud for facilitating access to publicly available CFHT $u$-band
data in the SXDS field, Victor Villar for providing data from his paper, and Robert
Yates for providing their best-fit projection for their local fundamental metallicity
relation. We also thank Brett A. Andrews for his insightful comments and discussion.
Our work is based in part on observations made with the NASA's {\it GALEX} mission.
We acknowledge support for this work from the GALEX Guest Investigator program under
NASA grants NNG09EG72I and NNX10AF04G.

{\it Facilities:} \facility{Mayall (NEWFIRM)}, \facility{Subaru (Suprime-Cam)},
\facility{Magellan (IMACS)}, \facility{GALEX}.


\LongTables
\begin{deluxetable*}{llllcccccc}
  \tabletypesize{\scriptsize}
  \tablewidth{0pt}
  \tablecolumns{10}
  \tablecaption{NEWFIRM Photometry of H$\alpha$-selected $z=0.8$ Galaxies}
  \tablehead{ 
    \colhead{} &\colhead{R.A.} &\colhead{Dec.} &\colhead{} &\colhead{} &\colhead{} &\colhead{} &\colhead{} &\colhead{} &\colhead{} \\		
    \colhead{ID} &\colhead{(J2000)} &\colhead{(J2000)} &\colhead{$z$} &\colhead{$m_{J}$} &\colhead{$m_{\rm NB118}$} &\colhead{Line flux} &\colhead{NB EW} &\colhead{\NII/H$\alpha$} &\colhead{$\log(L_{H\alpha})$} \\
 \colhead{(1)}	& \colhead{(2)}	&\colhead{(3)} &\colhead{(4)} &\colhead{(5)} &\colhead{(6)} &\colhead{(7)} &\colhead{(8)} &\colhead{(9)} &\colhead{(10)} }
  \startdata    
SXDSN-12615 &  2 17 17.7 & --4 45 18.9 &  0.796 &  21.46$\pm$0.07 &  20.79$\pm$0.05 &   19.6$\pm$2.0 &    56$\pm$17 &   0.27 &    41.660 \\
SXDSN-14848 &  2 18 09.7 & --4 42 25.1 &  0.805 &  22.96$\pm$0.24 &  21.83$\pm$0.09 &   10.6$\pm$1.7 &   125$\pm$94 &   0.40 &    41.364 \\
SXDSN-17153 &  2 17 29.1 & --4 47 11.4 &  0.797 &  20.56$\pm$0.02 &  19.90$\pm$0.01 &   41.5$\pm$1.5 &     48$\pm$5 &   0.38 &    41.953 \\
SXDSN-17287 &  2 18 11.5 & --4 47 01.0 &  0.805 &  22.75$\pm$0.22 &  21.99$\pm$0.13 &    7.2$\pm$1.9 &    68$\pm$62 &   0.35 &    41.213 \\
SXDSN-18372 &  2 18 11.6 & --4 46 17.3 &  0.804 &  21.52$\pm$0.07 &  20.94$\pm$0.05 &   14.8$\pm$1.9 &    43$\pm$16 &   0.31 &    41.537 \\
SXDSN-18643 &  2 18 20.5 & --4 46 06.6 &  0.804 &  21.18$\pm$0.04 &  20.71$\pm$0.03 &   15.3$\pm$1.7 &     32$\pm$9 &   0.41 &    41.519 \\
SXDSN-18689 &  2 18 05.2 & --4 46 05.9 &  0.805 &  21.67$\pm$0.06 &  21.03$\pm$0.04 &   15.4$\pm$1.4 &    53$\pm$15 &   0.37 &    41.536 \\
SXDSN-18825 &  2 17 24.3 & --4 45 55.8 &  0.797 &  22.70$\pm$0.16 &  22.22$\pm$0.12 &    3.5$\pm$1.6 &    29$\pm$33 &   0.31 &    40.908 \\
\enddata
  \label{table:genprop}
  \tablecomments{(1): ID from the NewH$\alpha$ survey catalog. 
    (2): Units of right ascension are hours, minutes, and seconds.
    (3): Units of declination are degrees, arcminutes, and arcseconds. 
    (4): IMACS spectroscopic redshift. 
    (5): $J$-band AB magnitude. The conversion from Vega to AB magnitudes is given by $m({\rm AB})-m({\rm Vega}) = 0.87$ \citep{ly11a}. 
    (6): Narrowband AB magnitude. The conversion from Vega to AB magnitudes is given by $m({\rm AB})-m({\rm Vega}) = 0.95$ \citep{ly11a}. 
    (7): Narrowband flux given in units of $10^{-17}$ erg s$^{-1}$ cm$^{-2}$. 
    (8): Narrowband EW (rest-frame) given in \AA. 
    (9): Estimated \NII/\Ha\ flux ratio, calculated using the narrowband EW (refer to Section \ref{sec:SFR_comp} for further details).
    Here, \NII\ refers to the doublet, 6548 and 6583 \AA. 
    (10): Log of observed H$\alpha$ luminosity given in units of erg s$^{-1}$. For both flux and luminosity, the amount of foreground
    extinction is negligible, so we excluded it in our calculations.}
\end{deluxetable*}

\clearpage

\begin{deluxetable*}{lcccccc}
  \tablewidth{0pt}
  \tablecolumns{7}
  \tablecaption{NewH$\alpha$ Emission-Line Fluxes from Magellan/IMACS}
  \tablehead{   
 	\colhead{ID}	&\colhead{[OII]\tablenotemark{a}}	&\colhead{[OIII]$\lambda$4959}	&\colhead{[OIII]$\lambda$5007}	&\colhead{H$\beta$}	&\colhead{H$\gamma$}		&\colhead{H$\delta$}  \\
	\colhead{(1)}	&\colhead{(2)}	&\colhead{(3)}	&\colhead{(4)}	&\colhead{(5)}	&\colhead{(6)}	&\colhead{(7)}}
   \startdata
SXDSN-12615 &  4.23$\pm$0.19 &    1.35$\pm$0.45 &    2.61$\pm$0.56 &    1.74$\pm$0.28 &  1.73$\pm$0.33 &  1.90$\pm$0.36 \\
SXDSN-14848 &  6.02$\pm$0.21 &    1.49$\pm$0.42 &    4.83$\pm$0.47 &    0.96$\pm$0.79 &  1.41$\pm$0.25 &  0.09$\pm$0.26 \\
SXDSN-17153 &  3.83$\pm$0.12 &    1.31$\pm$0.36 &           \ldots &    3.81$\pm$0.24 &  2.01$\pm$0.24 &  0.63$\pm$0.22 \\
SXDSN-17287 &  1.67$\pm$0.12 &    1.39$\pm$0.40 &    0.30$\pm$0.31 &    0.83$\pm$0.45 &  0.56$\pm$0.18 &  0.08$\pm$0.15 \\
SXDSN-18372 &  3.80$\pm$0.15 &    1.66$\pm$0.34 &    0.23$\pm$0.29 &    3.79$\pm$0.59 &  0.99$\pm$0.17 &  0.23$\pm$0.24 \\
SXDSN-18643 &  2.24$\pm$0.09 &    1.82$\pm$0.27 &    0.35$\pm$0.18 &    2.82$\pm$0.33 &  0.79$\pm$0.11 &         \ldots \\
SXDSN-18689 &  1.65$\pm$0.11 &    0.58$\pm$0.33 &    0.61$\pm$0.29 &    2.40$\pm$0.47 &  0.90$\pm$0.17 &  0.21$\pm$0.18 \\
SXDSN-18825 &  4.46$\pm$0.11 &    1.03$\pm$0.30 &    2.90$\pm$0.40 &    2.09$\pm$0.20 &  1.26$\pm$0.18 &  1.10$\pm$0.19 \\
\enddata
    \label{table:emline_flux}
  \tablecomments{All line fluxes are in units of $10^{-17}$ erg s$^{-1}$ cm$^{-2}$.}
  \tablenotetext{a}{The sum of the 3726 \AA\ and 3729 \AA\ lines.}
\end{deluxetable*}

\clearpage

\begin{deluxetable*}{lccccc}
  \tablewidth{0pt}
  \tablecolumns{6}
  \tablecaption{Stellar Mass, SFR, and AGN Classification}
  \tablehead{   
    \colhead{ID} &
    \colhead{Mass} &
    \colhead{SFR$_{\textrm{SED}}$} & 
    \colhead{SFR$_{\textrm{H}\alpha\textrm{,SED}}$} &
    \colhead{SFR$_{\textrm{H}\alpha\textrm{,Balmer}}$} &
    \colhead{AGN} \\
    \colhead{} &
    \colhead{log($M_{\odot}$)} &
    \colhead{log($M_{\odot}$ yr$^{-1}$)} &
    \colhead{log($M_{\odot}$ yr$^{-1}$)} &
    \colhead{log($M_{\odot}$ yr$^{-1}$)} &
    \colhead{} \\
    \colhead{(1)}	&
    \colhead{(2)}	&
    \colhead{(3)}	&
    \colhead{(4)}	&
    \colhead{(5)}	&
    \colhead{(6)}  }
  \startdata  
SXDSN-12615 & 10.23$\pm$0.11 &   0.76$\pm$0.10 &   0.68$\pm$0.29 & --1.75$\pm$0.75 & 0\\
SXDSN-14848 &         \ldots &          \ldots &   0.01$\pm$0.08 &          \ldots & 0\\
SXDSN-17153 & 11.15$\pm$0.06 &   1.88$\pm$0.05 &   1.73$\pm$0.19 &   0.27$\pm$0.38 & 0\\
SXDSN-17287 &  9.93$\pm$0.13 &   0.26$\pm$0.30 &   0.89$\pm$0.46 &          \ldots & 0\\
SXDSN-18372 & 10.01$\pm$0.08 &   0.70$\pm$0.09 &   0.81$\pm$0.35 &   1.79$\pm$0.64 & 0\\
SXDSN-18643 & 10.81          &   1.81          &   1.55$\pm$0.06 &   1.59$\pm$0.46 & 0\\
SXDSN-18689 & 10.48$\pm$0.12 &   0.77$\pm$0.33 &   1.32$\pm$0.42 &   0.79$\pm$0.74 & 0\\
SXDSN-18825 &  9.52$\pm$0.14 &   0.57$\pm$0.20 &   0.03$\pm$0.40 & --1.14$\pm$0.49 & 0\\
\enddata
  \label{table:msfr_agn}
  \tablecomments{Ellipses indicate weak measurements. (2): Stellar mass determinations are described in
    Section \ref{sec:sed}. (3): SFR based on SED fitting. (4): \Ha\ SFR, dereddened using the
    prescription of \citet{cha00} and SED results. (5): \Ha\ SFR, dereddened using the Balmer decrement.
    (6): ``1'' denotes AGN, using diagnostics described in Section \ref{sec:imacs}.}
\end{deluxetable*}


\begin{deluxetable*}{lcccccccc}
  \tablewidth{0pt}
  \tablecolumns{8}
  \tablecaption{NewH$\alpha$ SED-corrected Metallicities\label{table:z_sed_3sig}}
  \tablehead{ \colhead{ID} &\colhead{\citetalias{m91} upper} &\colhead{\citetalias{m91} lower} &
    \colhead{\citetalias{z94}} &\colhead{\citetalias{tre04}}
    &\colhead{N06} &\colhead{\citetalias{kk04} upper} &\colhead{\citetalias{kk04} lower} }
   \startdata
SXDSN-12615 & 8.661$\pm$0.120 & 7.986$\pm$0.179 & 8.768$\pm$0.132 & 8.713$\pm$0.119 &  8.700 &  8.828 &  8.177 \\
SXDSN-18825 & 8.568$\pm$0.079 & 8.149$\pm$0.092 & 8.672$\pm$0.097 & 8.635$\pm$0.068 &  8.601 &  8.734 &  8.302 \\
SXDSN-19822 & 8.944$\pm$0.002 & 7.362$\pm$0.006 & 9.102$\pm$0.003 & 9.020$\pm$0.003 &  9.066 &  9.049 &  7.680 \\
SXDSN-20554 & 8.956$\pm$0.026 & 7.408$\pm$0.069 & 9.122$\pm$0.032 & 9.041$\pm$0.033 &  9.094 &  9.066 &  7.701 \\
SXDSN-22048 & 8.295$\pm$0.183 & 8.487$\pm$0.202 & 8.358$\pm$0.256 & 8.400$\pm$0.172 & \ldots &  8.410 &  8.574 \\
SXDSN-22485 & 8.819$\pm$0.058 & 7.783$\pm$0.116 & 8.973$\pm$0.063 & 8.894$\pm$0.058 &  8.911 &  8.985 &  7.992 \\
SXDSN-23784 & 8.814$\pm$0.039 & 7.770$\pm$0.100 & 8.962$\pm$0.043 & 8.884$\pm$0.039 &  8.899 &  8.977 &  7.988 \\
SXDSN-23860 & 8.903$\pm$0.048 & 7.597$\pm$0.116 & 9.065$\pm$0.054 & 8.983$\pm$0.052 &  9.019 &  9.043 &  7.843 \\
\enddata
  \tablecomments{All metallicities are reported in form $Z=12+\log{({\rm O/H})}$. ``N06'' refers to \cite{n06}.}
\end{deluxetable*}

\clearpage

\begin{deluxetable*}{lccccccc}
  \tablewidth{0pt}
  \tablecolumns{8}
  \tablecaption{NewH$\alpha$ Balmer-corrected Metallicities\label{table:z_bal_3sig}}
  \tablehead{ \colhead{ID} &\colhead{\citetalias{m91} upper} &\colhead{\citetalias{m91} lower} &
    \colhead{\citetalias{z94}} &\colhead{\citetalias{tre04}}
    &\colhead{N06} &\colhead{\citetalias{kk04} upper} &\colhead{\citetalias{kk04} lower} }
   \startdata  
SXDSN-12615 & 9.077$\pm$0.009 & 6.776$\pm$0.080 & 9.301$\pm$0.017 & 9.218$\pm$0.015 & \ldots &  9.071 &  7.242 \\
SXDSN-18825 & 8.697$\pm$0.079 & 7.820$\pm$0.162 & 8.751$\pm$0.116 & 8.699$\pm$0.091 &  8.682 &  8.840 &  8.067 \\
SXDSN-19822 & 8.700$\pm$0.066 & 7.986$\pm$0.092 & 8.839$\pm$0.080 & 8.773$\pm$0.068 &  8.771 &  8.880 &  8.159 \\
SXDSN-20554 &          \ldots &          \ldots &          \ldots &          \ldots & \ldots & \ldots & \ldots \\
SXDSN-22048 & 9.068$\pm$0.021 & 7.048$\pm$0.136 & 9.276$\pm$0.041 & 9.196$\pm$0.040 & \ldots &  9.132 &  7.400 \\
SXDSN-22485 & 8.935$\pm$0.060 & 7.490$\pm$0.223 & 9.099$\pm$0.079 & 9.017$\pm$0.069 &  9.063 &  9.059 &  7.760 \\
SXDSN-23784 & 8.692$\pm$0.122 & 8.064$\pm$0.234 & 8.852$\pm$0.145 & 8.785$\pm$0.118 &  8.784 &  8.884 &  8.202 \\
SXDSN-23860 &          \ldots &          \ldots &          \ldots &          \ldots & \ldots & \ldots & \ldots \\
\enddata
  \tablecomments{All metallicities are reported in form $Z=12+\log{({\rm O/H})}$. ``N06'' refers to \cite{n06}.}
\end{deluxetable*}

\begin{deluxetable*}{lcccccc}
  \tablewidth{0pt}
  \tablecaption{Sample Cuts and Subsamples Used in Analysis} 
  \tablehead{
    \colhead{Sample Cut} &
    \colhead{$N$} &
    \colhead{$N$} &
    \colhead{SFR Comparison} &
    \colhead{\MSFR\ Relation} &
    \colhead{\MZ\ Relation} &
    \colhead{\MZS\ Relation} \\
    \colhead{} &
    \colhead{(total)} & 
    \colhead{(no AGN)} &
    \colhead{(Section~\ref{sec:SFR})} &
    \colhead{(Section~\ref{sec:mainseq})} &
    \colhead{(Section~\ref{sec:MZR})} &
    \colhead{(Section~\ref{sec:mzs_relation})} }
   \startdata  
     No restrictions                  & \Nspec & \Nnoagn &   &   &   &   \\
     Good SED fit ($\chi^{2}<10$)     & \Nthreenew & \Nthreeagn & X & X & X & X \\
     $u$-band photometry              & \Nfivenew & 160 & X &   &   &   \\
     Narrowband S/N $>3\sigma$        & 209 & 202 & X & X &   & X  \\
     Metallicity cut\tablenotemark{1} & 166 & 147 &   &   & X & X \\
     \hline 
     Total & & & \NSFRcompare & 188 & 136 & \Nthree 
   \enddata
   \tablenotetext{1}{The ``metallicity cut'' includes a $3\sigma$ restriction on \OII, \OII, and \Hb\
     (i.e., the $R_{23}(3\sigma)$ sample), as well as the removal of sources for which the calculated
     \citetalias{m91} upper-branch metallicity was lower than the \citetalias{m91} lower-branch metallicity.}
   \tablecomments{Note that AGN are removed for all four analyses listed. Also note that these
     sample cuts denote only the restrictions placed on datasets. Further restrictions (for linear or
     planar fits) are not listed here and are noted in the text and figures.}
   \label{table:subsample}
\end{deluxetable*}

\clearpage
\begin{deluxetable}{llllll}
  \tablewidth{0pt}
  \tablecolumns{6}
  \tablecaption{NewH$\alpha$ \MZ\ and \Mstar--SFR Relation Binned Points \label{table:bins}}
  \tablehead{ \multicolumn{3}{c}{\Mstar--SFR} & \multicolumn{3}{c}{\MZ} \\ 
      \colhead{\Mstar} & \colhead{SFR} & \colhead{$\sigma$} & \colhead{$M$} & \colhead{$Z$} & \colhead{$\sigma$} }
   \startdata  
   9.27 & 0.06 & 0.454 & 9.27 & 8.64 & 0.148 \\
   9.52 & 0.27 & 0.313 & 9.52 & 8.80 & 0.241 \\
   9.76 & 0.64 & 0.373 & 9.76 & 8.73 & 0.182 \\
  10.01 & 0.83 & 0.329 & 10.02 & 8.93 & 0.244 \\
  10.29 & 1.05 & 0.419 & 10.30 & 8.95 & 0.142 \\
  10.59 & 1.18 & 0.379 & 10.61 & 8.99 & 0.114 \\
  10.81 & 1.50 & 0.337 & 10.83 & 9.05 & 0.137 \\
  11.15 & 1.54 & 0.424 & 11.17 & 9.06 & 0.075 \\
  & & & 11.53 & 8.96 & 0.027 \\
  & & & 11.78 & 8.99 & 0.034 
   \enddata
   \label{table:binned}
\end{deluxetable}

\newcommand{\pa}{\phantom{1}}
\begin{deluxetable*}{lcccccccccc}
  \tablewidth{0pt}
  \tablecaption{Principal Component Analysis Planar Results}
  \tablehead{
    \colhead{Sample\tablenotemark{1}} & 
    \colhead{$N$} & 
    \colhead{Metallicity} &
    \colhead{$\alpha$} &
    \colhead{$\beta$} &
    \colhead{$\gamma$} &
    \colhead{$\delta$} &
    \colhead{$\sigma$} &
    \colhead{$P$(V1)} &
    \colhead{$P$(V2)} &
    \colhead{$P$(V1+V2)}\\
    \colhead{(1)} & 
    \colhead{(2)} & 
    \colhead{(3)} & 
    \colhead{(4)} &
    \colhead{(5)} &
    \colhead{(6)} &
    \colhead{(7)} &
    \colhead{(8)} &
    \colhead{(9)} &
    \colhead{(10)} &
    \colhead{(11)}}
   \startdata
   \cutinhead{\Ha-based SFRs}
   $R_{23}(3\sigma)$          &   119 & \citetalias{tre04} & --0.384$^{+0.03}_{-0.04}$ & 0.923$\pm$0.02        & 0.087$^{+0.04}_{-0.03}$ & 4.301$^{+0.57}_{-0.46}$   & 0.181 & 78.8$\pm$1.6 & 15.2$\pm$1.5 & 94.0$\pm$0.5\\[1mm]
   \Mstar\ $<10^{10}$ \Msun   & \pa79 & \citetalias{tre04} & --0.798$^{+0.13}_{-0.04}$ & 0.649$^{+0.11}_{-0.14}$ & 0.152$^{+0.05}_{-0.06}$ & --1.666$^{+1.90}_{-2.38}$ & 0.170 & 69.5$\pm$2.4 & 18.9$\pm$2.0 & 88.3$\pm$1.4\\[1mm]
   \Mstar\ $\ge10^{10}$ \Msun & \pa40 & \citetalias{tre04} & --0.322$^{+0.05}_{-0.04}$ & 0.947$\pm$0.02        & 0.103$\pm$0.04        & 5.403$^{+0.39}_{-0.78}$   & 0.159 & 73.3$\pm$3.2 & 21.3$\pm$3.1 & 94.5$\pm$1.0\\[1mm]
   $R_{23}(3\sigma)$          &   119 & \citetalias{m91}   & --0.329$^{+0.03}_{-0.04}$ & 0.946$^{+0.01}_{-0.02}$ & 0.063$^{+0.03}_{-0.04}$ & 5.006$^{+0.52}_{-0.41}$   & 0.170 & 79.5$\pm$1.6 & 15.1$\pm$1.5 & 94.6$\pm$0.5\\[1mm]
   \Mstar\ $<10^{10}$   \Msun & \pa79 & \citetalias{m91}   & --0.717$^{+0.17}_{-0.09}$ & 0.763$^{+0.12}_{-0.16}$ & 0.128$\pm$0.07        & --0.285$^{+2.85}_{-2.28}$ & 0.172 & 70.5$\pm$2.4 & 17.8$\pm$1.9 & 88.4$\pm$1.4\\[1mm]
   \Mstar\ $\ge10^{10}$ \Msun & \pa40 & \citetalias{m91}   & --0.270$^{+0.04}_{-0.03}$ & 0.964$^{+0.01}_{-0.02}$ & 0.093$\pm$0.04        & 5.875$^{+0.42}_{-0.52}$   & 0.136 & 74.5$\pm$3.2 & 21.2$\pm$3.1 & 95.8$\pm$0.7\\[1mm]
   $R_{23}(5\sigma)$          &    90 & \citetalias{tre04} & --0.446$\pm$0.05        & 0.884$^{+0.03}_{-0.02}$ & 0.127$\pm$0.04        & 3.444$^{+0.54}_{-0.68}$   & 0.175 & 78.7$\pm$1.9 & 15.4$\pm$1.7 & 94.1$\pm$0.6\\[1mm]
   \Mstar\ $<10^{10}$   \Msun & \pa61 & \citetalias{tre04} & --0.812$\pm$0.10        & 0.632$^{+0.13}_{-0.20}$ & 0.205$^{+0.06}_{-0.09}$ & --1.770$\pm$2.83        & 0.171 & 71.9$\pm$2.4 & 17.1$\pm$1.9 & 88.9$\pm$1.4\\
\enddata
   \tablecomments{See Section \ref{sec:PCA} for further details. PCA plane defined as
     $\alpha\log{(M_{\star}/M_{\sun})}+\beta[12+\log{(\rm O/H)}]+\gamma\log{({\rm SFR}/(M_{\sun}~{\rm yr}^{-1}))} = \delta$.
     The rms perpendicular to the PCA plane is provided in Column 8.}
   \tablenotetext{1}{The $R_{23}(3\sigma)$ and $R_{23}(5\sigma)$ samples are selected by
     requiring detections of \OII, \OIII, and \Hb\ at 3$\sigma$ and 5$\sigma$, respectively.}
   \label{table:PCA}
\end{deluxetable*}

\clearpage

\begin{deluxetable}{lccccc}
  \tablewidth{0pt}
  \tablecaption{$M=f(Z,{\rm SFR})$ Regression Planar Results}
  \tablehead{
    \colhead{Sample\tablenotemark{1}} & 
    \colhead{$N$} & 
    \colhead{Metallicity} &
    \colhead{$\beta_M$} &
    \colhead{$\gamma_M$} &
    \colhead{$\delta_M$}\\
    \colhead{(1)} & 
    \colhead{(2)} & 
    \colhead{(3)} & 
    \colhead{(4)} &
    \colhead{(5)} &
    \colhead{(6)}}
   \startdata  
   \cutinhead{\Ha-based SFRs}\\
   $R_{23}(3\sigma)$ & 119 & \citetalias{tre04} & 0.67$^{+0.13}_{-0.11}$ & 0.50$^{+0.04}_{-0.05}$ & 3.75$^{+1.02}_{-1.01}$\\[1mm]
   $R_{23}(3\sigma)$ & 119 & \citetalias{m91}   & 0.77$^{+0.14}_{-0.16}$ & 0.51$\pm$0.04        & 2.88$^{+1.35}_{-1.21}$\\[1mm]
   $R_{23}(5\sigma)$ &  90 & \citetalias{tre04} & 0.90$\pm$0.05        & 0.24$\pm$0.03        & 1.89$^{+0.49}_{-0.42}$\\[1mm]
   $R_{23}(5\sigma)$ &  90 & \citetalias{m91}   & 1.07$^{+0.08}_{-0.06}$ & 0.23$^{+0.02}_{-0.03}$ & 0.45$^{+0.57}_{-0.65}$\\[1mm]\hline
   \cutinhead{SED-based SFRs}\\
   $R_{23}(3\sigma)$ & 119 & \citetalias{tre04} & 0.25$^{+0.11}_{-0.13}$ & 0.59$^{+0.04}_{-0.05}$ & 7.16$^{+1.23}_{-1.07}$\\[1mm]
   $R_{23}(3\sigma)$ & 119 & \citetalias{m91}   & 0.30$^{+0.15}_{-0.18}$ & 0.59$\pm$0.04        & 6.76$^{+1.37}_{-1.20}$\\[1mm]
   $R_{23}(5\sigma)$ &  90 & \citetalias{tre04} & 0.69$^{+0.05}{-0.06}$ & 0.54$^{+0.04}_{-0.03}$ & 3.52$^{+0.56}_{-0.53}$\\[1mm]
   $R_{23}(5\sigma)$ &  90 & \citetalias{m91}   & 0.83$^{+0.07}_{-0.08}$ & 0.55$^{+0.04}_{-0.04}$ & 2.27$\pm$0.65\\
   \vspace{-3mm}
   \enddata
   \tablecomments{See Section~\ref{sec:reg} for further details. Regression plane defined as
     $\log{(M_{\star}/M_{\sun})} = \beta_M[12+\log{(\rm O/H)}]+\gamma_M\log{({\rm SFR}/(M_{\sun}~{\rm yr}^{-1}))} + \delta_M$.}
   \tablenotetext{1}{The $R_{23}(3\sigma)$ and $R_{23}(5\sigma)$ samples are selected by
     requiring detections of \OII, \OIII, and \Hb\ at 3$\sigma$ and 5$\sigma$, respectively.}
   \label{table:RegM}
\end{deluxetable}

\begin{deluxetable*}{lcccrc}
  \tablewidth{0pt}
  \tablecaption{$Z=f(M,{\rm SFR})$ Regression Planar Results}
  \tablehead{
    \colhead{Sample\tablenotemark{1}} & 
    \colhead{$N$} & 
    \colhead{Metallicity} &
    \colhead{$\alpha_Z$} &
    \colhead{$\gamma_Z$} &
    \colhead{$\delta_Z$}\\
    \colhead{(1)} & 
    \colhead{(2)} & 
    \colhead{(3)} & 
    \colhead{(4)} &
    \colhead{(5)} &
    \colhead{(6)}}
   \startdata  
   \cutinhead{\Ha-based SFRs}\\
   $R_{23}(3\sigma)$ & 119 & \citetalias{tre04} & 0.23$\pm$0.02        & 0.01$^{+0.02}_{-0.03}$   & 6.61$^{+0.18}_{-0.12}$\\[1mm]
   $R_{23}(3\sigma)$ & 119 & \citetalias{m91}   & 0.19$\pm$0.02        & 0.00$\pm$0.02          & 6.97$^{+0.18}_{-0.15}$\\[1mm]
   $R_{23}(5\sigma)$ &  90 & \citetalias{tre04} & 0.26$^{+0.03}_{-0.02}$ & 0.00$^{+0.03}_{-0.02}$   & 6.32$^{+0.25}_{-0.24}$\\[1mm]
   $R_{23}(5\sigma)$ &  90 & \citetalias{m91}   & 0.22$\pm$0.02        & --0.01$\pm$0.02        & 6.68$^{+0.19}_{-0.20}$\\[1mm]\hline
   \cutinhead{SED-based SFRs}\\
   $R_{23}(3\sigma)$ & 119 & \citetalias{tre04} & 0.25$\pm$0.02        & --0.03$^{+0.03}_{-0.02}$ & 6.39$^{+0.22}_{-0.19}$\\[1mm]
   $R_{23}(3\sigma)$ & 119 & \citetalias{m91}   & 0.21$\pm$0.02        & --0.03$\pm$0.02        & 6.81$^{+0.19}_{-0.14}$\\
\enddata
   \tablecomments{See Section~\ref{sec:reg} for further details. Regression plane defined as
     $12+\log{(\rm O/H)} = \alpha_Z \log{(M_{\star}/M_{\sun})} +\gamma_Z\log{({\rm SFR}/(M_{\sun}~{\rm yr}^{-1}))} + \delta_Z$.}
   \tablenotetext{1}{The $R_{23}(3\sigma)$ and $R_{23}(5\sigma)$ samples are selected by
     requiring detections of \OII, \OIII, and \Hb\ at 3$\sigma$ and 5$\sigma$, respectively.}
   \label{table:RegZ}
\end{deluxetable*}

\end{document}